\documentclass[a4paper,11pt]{article}
\pdfoutput=1 % if your are submitting a pdflatex (i.e. if you have
\usepackage{jheppub} % for details on the use of the package, please
\usepackage{amsmath,amssymb,amsthm,amscd,graphicx}
\usepackage{psfrag}
\input epsf.sty

\usepackage{color}

\theoremstyle{definition}

\usepackage{titlesec}

\titleformat{\paragraph}
{\normalfont\normalsize\bfseries}{\theparagraph}{1em}{}
\titlespacing*{\paragraph}
{0pt}{3.25ex plus 1ex minus .2ex}{1.5ex plus .2ex}

\newcommand{\CA}{{\cal A}}

\newcommand{\CF}{{\cal F}}

\newcommand{\CN}{{\cal N}}
\newcommand{\CO}{{\cal O}}

\newcommand{\CZ}{{\cal Z}}

\newcommand{\fad}{\operatorname{\Phi}_{\mathsf{b}}}

\def\IN{{\mathbb N}}
\def\IZ{{\mathbb Z}}
\def\IR{{\mathbb R}}
\def\IC{{\mathbb C}}
\def\IQ{{\mathbb Q}}

\def\IP{{\mathbb P}}

\def\IF{{\mathbb F}}

\newcommand{\re}{{\rm e}}
\newcommand{\ri}{{\rm i}}
\newcommand{\rd}{{\rm d}}

\newcommand{\be}{\begin{equation}}
\newcommand{\ee}{\end{equation}}
\newcommand{\ba}{\begin{aligned}}
\newcommand{\ea}{\end{aligned}}
\newcommand{\ben}{\begin{eqnarray}\displaystyle}
\newcommand{\een}{\end{eqnarray}}

\newdimen\tableauside\tableauside=1.0ex
\newdimen\tableaurule\tableaurule=0.4pt
\newdimen\tableaustep
\def\phantomhrule#1{\hbox{\vbox to0pt{\hrule height\tableaurule width#1\vss}}}
\def\phantomvrule#1{\vbox{\hbox to0pt{\vrule width\tableaurule height#1\hss}}}
\def\sqr{\vbox{%
  \phantomhrule\tableaustep
  \hbox{\phantomvrule\tableaustep\kern\tableaustep\phantomvrule\tableaustep}%
  \hbox{\vbox{\phantomhrule\tableauside}\kern-\tableaurule}}}
\def\squares#1{\hbox{\count0=#1\noindent\loop\sqr
  \advance\count0 by-1 \ifnum\count0>0\repeat}}
\def\tableau#1{\vcenter{\offinterlineskip
  \tableaustep=\tableauside\advance\tableaustep by-\tableaurule
  \kern\normallineskip\hbox
    {\kern\normallineskip\vbox
      {\gettableau#1 0 }%
     \kern\normallineskip\kern\tableaurule}%
  \kern\normallineskip\kern\tableaurule}}
\def\gettableau#1{\ifnum#1=0\let\next=\null\else
\squares{#1}\let\next=\gettableau\fi\next}

\newcommand{\figref}[1]{Fig.~\protect\ref{#1}}
\newcommand*{\tabref}[1]{\tablename~\ref{#1}}
\title{\boldmath Quantum curves and $q$--deformed Painlev\'e equations}
\author{ Giulio Bonelli$^a$, Alba Grassi$^{b,c}$ and Alessandro Tanzini$^a$}

\affiliation{
$^a$ International School of Advanced Studies (SISSA), \\
via Bonomea 265, 34136 Trieste, Italy and INFN, Sezione di Trieste\\
\\
$^b$International Center for Theoretical Physics,\\
 ICTP, Strada Costiera 11, Trieste 34151, Italy \\
 \\
$^c$ Simons Center for Geometry and Physics,\\
 SUNY, Stony Brook, NY, 1194-3636, USA \\}

\emailAdd{bonelli@sissa.it, agrassi@scgp.stonybrook.edu, tanzini@sissa.it}
\preprint{
\begin{flushright}
SISSA  51/2017/FISI-MATE \\
\end{flushright}
}

\abstract{We propose that the grand canonical topological string partition functions satisfy finite-difference equations in the closed string moduli. In the case of genus one mirror curve these are conjectured to be 
the q-difference Painlev\'e equations as in Sakai's classification. More precisely, we propose that 
the tau-functions of q-Painlev\'e equations are related to the grand canonical topological string partition functions on the corresponding geometry. In the toric cases we use topological string/spectral theory duality to give a Fredholm determinant representation for  the above tau-functions in terms of the  underlying  quantum mirror curve. As a consequence, the zeroes of the tau-functions compute the exact spectrum of the associated quantum integrable systems. We provide details of this construction for the local $\mathbb{P}^1\times \mathbb{P}^1$ case, which is related to q-difference Painlev\'e with affine $A_1$ symmetry, to $SU(2)$ Super Yang-Mills in five dimensions and to relativistic Toda system. }

\begin{document}
\maketitle

\flushbottom

\section{Introduction and Summary}

During the last decades an intriguing relationship has been observed between the low-energy dynamics of $\mathcal{N}=2$ four-dimensional gauge theories and integrable systems \cite{sw1,sw2,Martinec:1995by,Gorsky:1995zq}.
The use of localization techniques in the supersymmetric path integral considerably clarified and widened this relation in many directions \cite{no2, ns}. In this context, a precise connection between  supersymmetric partition
functions and tau functions of isomonodromic deformation problems associated to the Seiberg-Witten geometry has been established  leading to the Painlev\'e/ $SU(2)$ gauge correspondence \cite{ilt1,ilt,gil1,ilte, bes,gil,blmst,Nagoya:2016aa,lnew,Gavrylenko:2017lqz}. 
More precisely, it was found that  $\tau$ functions of differential Painlev\'e  equations are computed by the Nekrasov--Okounkov (NO) \cite{no2} partition functions of four dimensional $SU(2)$, $\CN=2$ gauge theories in the self--dual phase of the $\Omega$ background.  The specific matter content of the gauge theory determines the type of Painlev\'e equation (see Table 2 in \cite{blmst} for the precise relation).  Moreover the long/short distance expansions of Painlev\'e equations are in correspondence with the duality frames in the gauge theory   \cite{blmst,bgt}.

On the other hand, by resorting to the geometric engineering of gauge theories via topological strings  \cite{kkv}, it has been possible to show in some cases \cite{bgt,bgt2} that these tau functions are Fredholm (or spectral) determinants of quantum operators  arising in a suitable limit of the non-perturbative topological string formulation of \cite{ghm,cgm2}. 
This embedding into topological strings has allowed to take a first step through the generalisation  of the 
Painlev\'e/ SU(2) gauge correspondence to the higher rank case providing explicit Fredholm determinant representation for the $SU(N)$ theories  \cite{bgt2}, see also \cite{gavry}. 
Furthermore from the string theory viewpoint it is also natural to consider the five dimensional version of this correspondence as pointed out in  \cite{bsu,blmst,bgt}  and further   studied in  \cite{Jimbo:2017aa,Mironov:2017sqp}\footnote{   In \cite{Mironov:2017sqp}, based on \cite{Mironov:2017gja, Mironov:2017lgl}, a different type of  finite dimensional determinant was considered to compute $\tau$ functions.}. 
 On the Painlev\'e side this corresponds to a lift from differential to difference equations. The latter arises as a $q$--deformation of   Painlev\'e  equations and we refer to them as  $q$-Painlev\'e  ($q$--P) equations, see \cite{Kajiwara:2015aa, Grammaticos2004}  for a  review and a  list of references.  
 On the gauge theory 
 side instead this corresponds to a lift from  four  dimensional $SU(2)$, $\CN=2$ gauge theories to topological string on local CY manifolds.  In particular 
  one uses  topological string theory to compute tau-functions of $q$-Painlev\'e equations as defined and studied in \cite{sakai,Kajiwara:2015aa}. 

\begin{figure}[h!] \begin{center}
 {\includegraphics[scale=0.3]{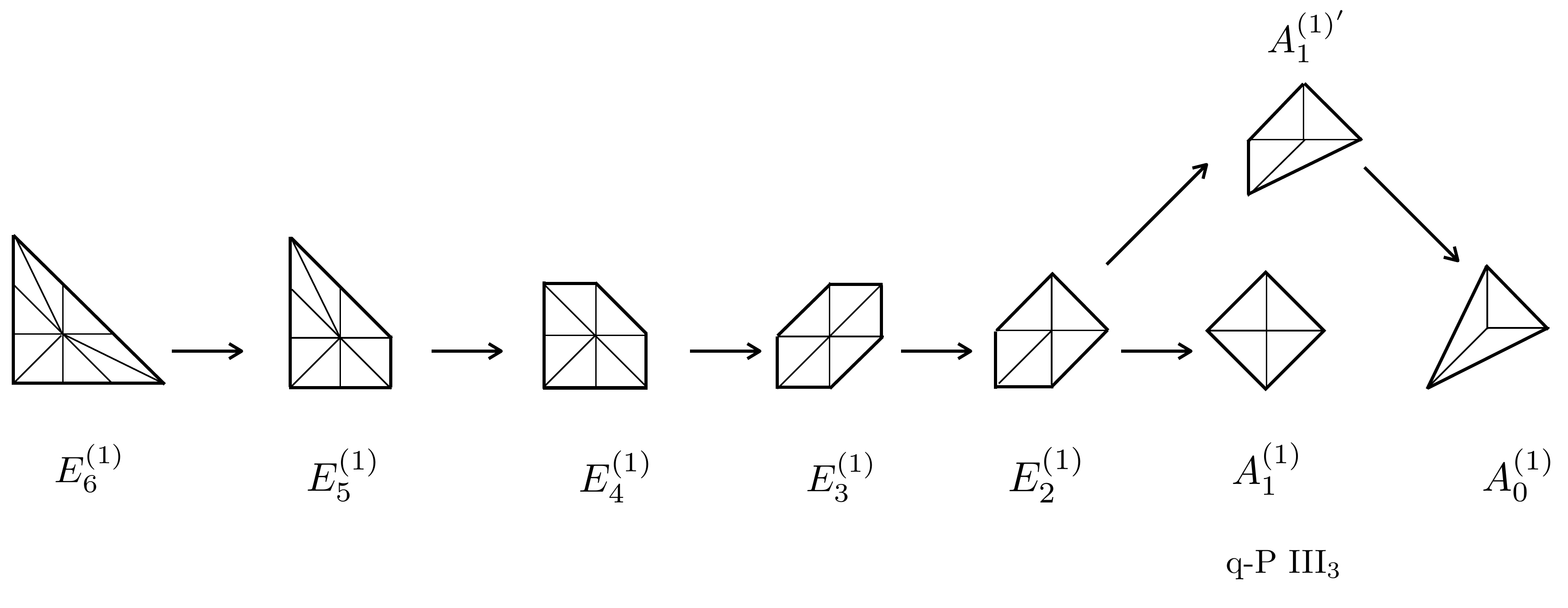}}
\caption{
 The letters  $E_{\cdots}^{\cdots}$ and $A_{\cdots}^{\cdots}$ refers to symmetry type classification of $q$-Painlev\'e  according to \cite{sakai,Kajiwara:2015aa}.  
The Newton polygons on the upper line are the ones associated to the $q$-Painlev\'e according to \cite{yamadaameba,Ormerod:2014aa}.  This correspondence is not always  unique as discussed in Section \ref{qpnp}.  }\label{p-toric}
\end{center}
\end{figure}

Although we expect the correspondence between topological strings and Painlev\'e to hold in general for all Painlev\'e equations
in Sakai's classification \cite{sakai} (see \tabref{tab1}),  
in this work we will focus on the ones with a {\it toric} topological string realization (see \figref{p-toric}).
%%%%
\begin{table}[ht]
\begin{center}
 \begin{tabular}{ll}
	Painlev\'e   type   & Physical theory    \\ \hline
	Elliptic           &  E-strings\\
	Multiplicative  \quad  & Topological string on local del Pezzo's.\\
	Additive         &  4-dimensional $SU(2)$ gauge theories\\
	     \end{tabular}
        \caption{  On the left: classification of  Painlev\'e equations according to \cite{sakai,Kajiwara:2015aa}. The additive type correspond to the standard differential Painlev\'e  equations
         plus the three finite additive cases corresponding to Minahan-Nemeshanski four-dimensional gauge theories \cite{Minahan:1996cj}.
        The multiplicative cases correspond to q-difference Painlev\'e (see \cite{Kajiwara:2015aa, sakai} for more details). On the right: physical theory that we expect to compute the tau functions of  Painlev\'e equations. In the multiplicative case one has to consider  blow up of del Pezzo up to  8 blow up, while the Elliptic Painlev\'e makes contact with {${1\over 2} K3$}.  The analogy between Sakai's scheme for Painlev\'e equations  and the geometry underling the above physical theories was originally suggested in \cite{Mizoguchi:2002kg}.    }  \label{tab1}
\end{center}
          \end{table}
%
%%%%
We will give a prescription to construct a Fredholm determinant representation of such tau-functions starting from the geometrical formulation of $q$-Painlev\'e
presented in \cite{yamadaameba,Ormerod:2014aa}. As we will outline in section \ref{gen},  the first step consists in  associating a Newton polygon to these difference equations as illustrated for instance in  \figref{p-toric}.
Such a polygon can then be related to the toric diagram of a corresponding  Calabi--Yau (CY) manifold. We will conjecture that the Fredholm determinant of  the operator arising in the quantization of its mirror curve  computes the  tau-function of the corresponding $q$-difference Painlev\'e equation.
 As a consequence  the zeros of such tau-function  compute the exact spectrum of the integrable systems associated to the underlying mirror curve \cite{Goncharov:2011hp,Fock:2014ifa}. In this way we  also provide a concrete link between $q$-Painlev\'e equations and the  topological string/spectral theory (TS/ST) duality \cite{ghm}. 
 We remark that from the topological string viewpoint $q$-Painlev\'e equations control the dependence of the partition function on the {\it closed} string moduli. This is in line with 
the expectation of exact quantum background independence which should be fulfilled by a non-perturbative formulation of topological string theory \cite{witten-bi}\footnote{We would like to thank Marcos Mari\~no  for a discussion on this point.}. 
 
The structure of the paper is the following. In Sect. \ref{gen} we outline the general features of the correspondence between topological strings, spectral theory and $q$-Painlev\'e.
In the subsequent sections we work out explicitly the  example of the $q$-difference Painlev\'e $\rm III_3$ \footnote{In Sakai's classification \cite{sakai} this corresponds to  
surface type $A_7^{(1)'} $  and symmetry type {  $A_1^{(1)}$}, where the superscript $(1)$ stands for affine extension of the Dynkin algebra. {Notice that this is not the unique q-P equation leading to differential Painlev\'e $\rm III_3$ in the continuous limit. Indeed, as pointed out for instance in \cite{rgi}, also the  q-P equation with surface type $A_7^{(1)} $  and symmetry type {  $A_1^{(1)'}$} makes contact with  Painlev\'e $\rm III_3$. This is perfectly consistent with the picture developed in this paper since the corresponding Newton polygon is identified with local $\IF_1$ (see \figref{p-toric}). By the geometric engineering construction \cite{kkv} we know that topological string theory on both $\IF_1$  and $\IP^1\times \IP^1$  reduces to pure $SU(2)$ theory in four dimensions. However in this work we denote by q-P  $\rm III_3$ only the one associated to surface type $A_7^{(1)'} $  and symmetry type {  $A_1^{(1)}$}. } } which makes contact with topological strings on local $\IP^1\times \IP^1$.
  In this example the TS/ST duality \cite{ghm} states that 
  \be \label{introts}\Xi^{\rm TS}_{\IP^1\times \IP^1}(\kappa,\xi, \hbar)=\det \left(1+\kappa \rho_{\IP^1 \times \IP^1}\right) , \ee
  where 
  \be\label{grandcan}   \Xi^{\rm TS}_{\IP^1\times \IP^1}(\kappa,\xi, \hbar)=  \sum_{n \in \IZ}\re^{\mathsf J_{\IP^1\times \IP^1}(\mu+2 \pi \ri n, { \xi}, \hbar)}\quad \kappa=\re^{\mu} \ee
  is the grand canonical topological string partition function,  $\mathsf J_{\IP^1\times \IP^1}$ is the topological string grand potential (see appendix \ref{def}) and  $ \rho_{\IP^1 \times \IP^1}$ is the operator arising in the quantisation of the mirror curve to local $\IP^1 \times \IP^1$ (see equation \eqref{rhop1p1}).  In section \ref{s3} we show that   \eqref{grandcan} satisfies the  $q$-difference Painlev\'e $\rm III_3$ equation in the $\tau$ form. As a consequence this provides a conjectural Fredholm  determinant solution for the corresponding $\tau$ function whose explicit expression is given on the r.h.s.~of \eqref{introts}.  As shown in \cite{bgt} it exists a suitable limit in which 
  \be \label{introdet}\det \left(1+\kappa \rho_{\IP^1 \times \IP^1}\right)\ee 
  reduces to a well known the determinant computing the tau function of the standard Painlev\'e $\rm III_3$   \cite{zamo,wu1}. From that perspective our result can be viewed as a   generalisation of  \cite{zamo,wu1} for the $q$-deformed  Painlev\'e equations.

In section \ref{s4} we discuss the $q$-deformed algebraic solution associated to  such a Fredholm determinant representation, while in section \ref{s5} and \ref{s6} we test by explicit computations that the r.h.s.~of \eqref{introts} indeed fulfils the $q$-difference Painlev\'e $\rm III_3$ equation in the $\tau$ form. Moreover, in section \ref{s6} we connect our results with ABJ theory by relating $q$-Painlev\'e equations to Wronskian-like relations of \cite{ghmabjm}. Section \ref{s7} is devoted to conclusions and open problems. In the appendices we collect some technical results and definitions.

\section{Generalities} \label{gen}
In this work we propose that  spectral determinants of  operators arising in the quantization of  mirror curve to CY manifolds compute $\tau$ functions of $q$-deformed Painlev\'e equations when some particular initial conditions are imposed. We start by reviewing some results which are relevant for this proposal. 

\subsection{Topological string and spectral theory}
\label{s2}
In this section we review the results of \cite{ghm} in a form which is suitable for the propose of this work.
Let us consider a toric CY $X$ with genus one mirror curve \footnote{For the higher genus generalisation see \cite{cgm2}.}.  By following \cite{hkp,hkrs} the complex moduli of the mirror curve to $X$ are divided in two classes of parameters: one "true" modulus, denoted by $\kappa$, and  $r_X$ mass parameters, denoted by \be \label{mass} {\boldsymbol m_{X}}=\left\{m_{X}^{(1)}, \cdots, m_{X}^{(r_X)}\right\}.\ee
We introduce the rescaled mass parameters   $\bf m$ as  \cite{kmz,bgt}
\be \label{rescaled} \log m^ {(i)}= {2\pi\over \hbar} \log \left(m_{X}^{(i)}\right),\ee
as well as
\be { \xi^{(i)}}=\log {m^{(i)}}.\ee 
The  Newton polygon  of $X$ is defined as the convex hull of a set of two-dimensional vectors  \cite{bat,ckyz,hv,kkv}
\be \vec{\nu}^{(i)}=\left(\nu_1^{(i)}, \nu_2^{(i)}\right), \quad i=1, \cdots,k , \ee
from which one reads the  mirror curve to $X$  as 
\be \label{mirr} \qquad  \sum_{i=1}^{k} \re^{\nu_1^{(i)}x+\nu_2^{(i)}p+f_i ({\bf m}_X)} +\kappa=0, \quad x, p \in \IC \ee
where $f_i$ is a function of the mass parameters.
\begin{figure}[h!] \begin{center}
 {\includegraphics[scale=0.4]{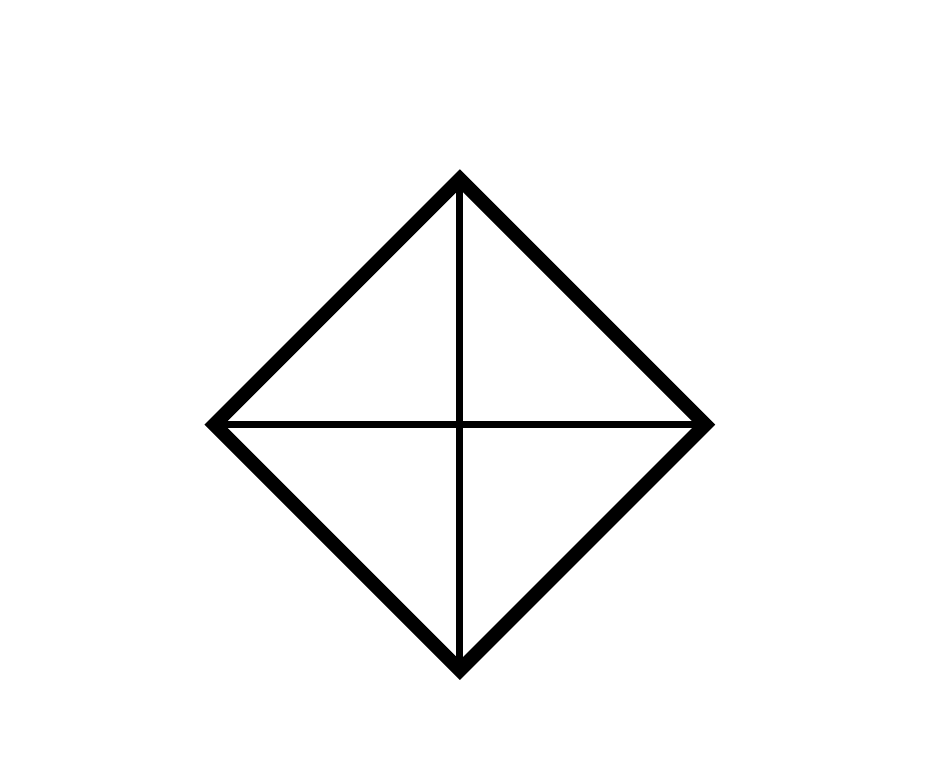}}
\caption{  Newton polygon  of local $\IP^1\times \IP^1$  }\label{toricp1}
\end{center}
\end{figure} 
For instance  when $X$ is the canonical bundle over $\IP^1\times \IP^1$ its Newton polygon  is shown in \figref{toricp1}. 
The corresponding  vectors are
\be \vec{\nu}^{(1)}=\{1,0\}, \quad \vec{\nu}^{(2)}=\{-1,0\}, \quad  \vec{\nu}^{(3)}=\{0,1\}, \quad \vec{\nu}^{(4)}=\{0,-1\}.\ee
Therefore the mirror curve reads
\be \label{mirror} \re^x +m_{\IP^1\times \IP^1}\re^{-x}+\re^{p}+\re^{-p}+\kappa=0. \ee
We introduce the quantum operators $ \mathsf x,\mathsf p $ such that
\be [\mathsf x, \mathsf p]=\ri \hbar, \ee
and we promote the mirror  curve  \eqref{mirr} to a quantum operator by using Weyl quantization. The resulting operator is
\be \label{op} {\mathsf O}_{ X}=\sum_{i=1}^{k}\re^{f_i ({\bf m}_X)} \re^{\nu_1^{(i)}\mathsf x+\nu_2^{(i)} \mathsf p}. \ee
 An explicit list of these operators can be found in Table 1 of \cite{ghm}.
 It was conjectured in \cite{ghm}, and later proved in \cite{kama,Laptev:2015loa} for many geometries, that the inverse 
 \be\label{rhoo}  { \rho}_{X}={\mathsf O}_X^{-1}\ee
 is a self-adjoint, positive\footnote{Provided some positivity constraints are imposed on  the mass parameters  and  $\hbar$.} and  trace class operator acting on $L^2(\IR)$. Therefore its spectral, or  Fredholm,  determinant 
 \be  \label{Fred}\Xi_X^{\rm ST}(\kappa, {\boldsymbol \xi},\hbar)= \det\left(1+\kappa \rho_X\right)\ee
is analytic in $\kappa$.   The  operator \eqref{rhoo} is related to the Hamiltonian of a corresponding integrable system \cite{Goncharov:2011hp,Fock:2014ifa}.
The conjecture of  \cite{ghm} states that the spectral determinant \eqref{Fred} of  operators arising in the quantization of the mirror curves to $X$ is computed by (refined) topological string theory on $X$. This has led to a new and exact relation between the spectral theory of quantum mechanical operators and enumerative geometry/topological string theory. We refer to it as TS/ST duality. 
Let us define the grand canonical topological string partition function on $X$ as 
\be \label{ts} \Xi_X^{\rm TS}(\kappa, {\boldsymbol \xi},\hbar)=\sum_{n \in \IZ}\re^{\mathsf J_X(\mu+2 \pi \ri n, {\boldsymbol \xi}, \hbar)} , \qquad \kappa=\re^{\mu}, \qquad \hbar\in \IR_+\ee
where we denote by  $\mathsf J_X$  the topological string grand potential studied in \cite{mp,hmo2,cm,hmo3,hmmo}.  By following \cite{cgm2,Marino:2015nla} we write $\mathsf J_X$ as
\be \label{jts}\mathsf J_X(\mu,{\boldsymbol \xi}, \hbar)=\mathsf J_X^{\rm WS}(\mu,{\boldsymbol \xi}, \hbar)+ \mathsf J_X^{\rm WKB}(\mu, {\boldsymbol \xi},\hbar), \ee
where $ \mathsf J_X^{\rm WS}$ is expressed in terms of unrefined topological string while $ \mathsf J_X^{\rm WKB}$ is determined by the Nekrasov--Shatashvili (NS) limit of the refined topological string.  The precise definitions can be found in appendix \ref{def}. 
We would like to stress that, even though both $\mathsf J_X^{\rm WKB}$ and $\mathsf J_X^{\rm WS}$ have a dense set of poles on the  real $\hbar$ axis, their sum is  well defined and free of poles. This is the so--called HMO cancellation mechanism  \cite{hmo2} which was first discovered in the context of ABJM theory and has played an important role in the TS/ST duality presented in \cite{ghm}. One can also  write \eqref{ts} as \cite{ghm}
\be \label{qtheta} \Xi_X^{\rm TS}(\kappa, {\boldsymbol \xi},\hbar)=\re^{\mathsf J_X(\mu, {\boldsymbol \xi}, \hbar)} \Theta_X(\mu, {\boldsymbol \xi}, \hbar) ,\ee
where $\Theta_X$ defines a quantum theta function which,  for some specific values of $\hbar$, becomes a conventional theta function. This happen for instance when $\hbar=2 \pi/m$ for $m \in \IN$  as explained in \cite{cgm, ghm, huang1606}.
The conjecture \cite{ghm} states that 
\be \label{tsst} \Xi_X^{\rm TS}(\kappa, {\boldsymbol \xi},\hbar)=\Xi_X^{\rm ST}(\kappa, {\boldsymbol \xi},\hbar).\ee
Even though we still do not have a rigorous mathematical proof of \eqref{tsst},  many aspect and consequences of this proposal have been successfully tested and proved in severals examples both numerically and analytically
\cite{ghm, kama,mz,kmz,gkmr,cgm2, oz,wzh,hm,hw,fhma,bgt,bgt2,ag,hkt2,mz2,huang1606, cgum,ggu,Sciarappa:2017hds,Couso-Santamaria:2016vwq, sug,Marino:2017gyg,Hatsuda:2017zwn, ghkk}. 

 Originally the above construction was formulated only for real values of $\hbar$. Nevertheless it was found in \cite{gmprogress} that when the underling geometry can be used to engineer gauge theories one can easily extend some aspects of \cite{ghm} to generic complex value of  $\hbar$. However in this work we will focus on the real case. 

\subsubsection{Self--dual point }\label{sdpoint}
It was  pointed out in \cite{cgm,ghm} that there is a particular value of $\hbar$ where the grand potential simplify drastically, this occurs at $\hbar=2\pi$ and we refer to it as self--dual point. At this point $\mathsf J_X$ is determined   only by genus zero and genus one free energies \eqref{f01}, \eqref{fgns}. Therefore at this point we can express the spectral determinant  \eqref{Fred} in closed form  in term of hypergeometric, Meijer  and theta functions. Modular properties of  the spectral determinant around this point have been discussed in \cite{ag}.  The explicit expression for  $ \mathsf J_X(\mu, {\boldsymbol \xi}, 2\pi)$ and $\Xi_X^{\rm TS}(\kappa, {\boldsymbol \xi}, 2\pi)$ for generic $X$ at the self--dual point can be found in section  3.4 of \cite{ghm} or section of 3.2 of  \cite{cgm2}.
As explained in these references, the genus one free energy appears as an overall multiplication factor, while the dependence on the genus zero free energy and its derivatives is non trivial. In this section we re-write the details  only for two explicit examples, which we will use later. These cases have been worked out in \cite{cgm}.

We first consider local $\IP^1\times \IP^1$ with $\xi=2 \pi \ri$ \footnote{This correspond to ABJM theory with level  $k=2$ \cite{cgm}. In the ABJM  context the self--dual point correspond to an enhancement of the supersymmetry from $\CN=6$ to $\CN=8$.  }. Then we have \cite{cgm} \footnote{For sake of notation we will simply denote $\Xi^{\rm TS}(\kappa, \xi, \hbar)$ instead of $\Xi^{\rm TS}_{\IP^1\times \IP^1}(\kappa, \xi, \hbar)$ }
\be 
\label{std2} \ba 
\Xi^{\rm TS}(\kappa, 2 \pi, 2 \pi)= & \exp \left[{\CA(\ri \kappa)}\right] \vartheta_3( \xi(\ri \kappa),  \tau(\ri \kappa)),
\ea\ee
where \be \CA(\kappa)= {\log \kappa \over 4}+\CF_1+F_1^{\rm NS} -{1\over  \pi^2}  \left( \CF_0(\lambda)-\lambda \partial_{\lambda}\CF_0(\lambda)+{\lambda^2 \over 2} 
\partial^2_{\lambda}\CF_0(\lambda)\right).\ee
We denote by $F_1^{\rm NS}$  the NS genus one free energy \eqref{fgns} which in the present example  reads
\be F^{\rm NS}_1= -\frac{1}{24} \log \left(16+\kappa ^2\right)-\frac{\log (\kappa )}{12}.\ee
Moreover  $\CF_0$ and $\CF_1$  are the genus zero and genus one free energies in the orbifold frame. These can be obtained from $F_0$ and $F_1$ in \eqref{f01} by using modular transformation and analytic continuation as explained in \cite{abk, mpabjm,dmp}. More precisely we have
\be 
\label{standard}
\ba
 \partial_\lambda \CF_0 (\lambda)&= { \kappa \over 4 } G^{2,3}_{3,3} \left( \begin{array}{ccc} {1\over 2}, & {1\over 2},& {1\over 2} \\ 0, & 0,&-{1\over 2} \end{array} \biggl| -{\kappa^2\over 16}\right)+ { \pi^2 \ri \kappa\over 2}   {~}_3F_2\left(\frac{1}{2},\frac{1}{2},\frac{1}{2};1,\frac{3}{2};-\frac{\kappa^2
   }{16}\right),  \ea \ee
   with
\be
\CF_0(\lambda)=-4 \pi ^2 \lambda ^2 \left(\log (2 \pi  \lambda )-\frac{3}{2}-\log (4)\right)+\cdots
\ee   and we denote by $\lambda$  the quantum  K\"ahler  parameter at $\hbar=2\pi$ in the orbifold frame namely
   \be \label{lamb}\lambda={\kappa \over 8 \pi}  {~}_3F_2\left(\frac{1}{2},\frac{1}{2},\frac{1}{2};1,\frac{3}{2};-\frac{\kappa^2
   }{16}\right). 
\ee
Our convention for the hypergeometric functions  $G^{2,3}_{3,3}(\cdot)$ and ${}_3F_2(\cdot)$ in \eqref{standard} and \eqref{lamb} are as in \cite{cgm}.
As for the genus one free energy we have
\be
\label{F1st} \CF_1=- \log \eta\left( 2  \tau\right)-{1\over 2}\log 2,
 \ee
where $\eta$ is the Dedekind eta function and we used
\be \label{taudef} \tau(\kappa)=-{1 \over 8 \pi^3 \ri} \partial_\lambda^2 \CF_0 (\lambda)=-\frac{1}{2}+\frac{\ri K\left(\frac{\kappa ^2}{16}+1\right)}{2 K\left(-\frac{\kappa ^2}{16}\right)},\ee
where $K(\kappa^2)$ is the elliptic integral of first kind. 
In \eqref{std2} we denote by $ \vartheta_3$  the  Jacobi theta function
\be \vartheta_3(v,\tau)= \sum_{n \in \IZ}  \exp \left[ \pi \ri n^2 \tau + 2 \pi \ri n v\right]\ee
and we define
\be 
 \xi(\kappa)= {\ri \over 4 \pi^3} \left( \lambda \partial_\lambda^2 \CF_0 (\lambda) - \partial_\lambda \CF_0 (\lambda)\right).
\ee
Noticed that many of the quantities defined above have branch cuts, however according to \cite{cgm} these should cancel and the final answer \eqref{std2} is analytic in $\kappa$.

Likewise for local $\IP^1\times \IP^1$, $\hbar=2 \pi$ and $\xi=0$ \footnote{This correspond to ABJ theory with level  $k=2$ and gauge group $U(N)\times U(N+1)$\cite{cgm}.} we have \cite{cgm} 
 \be 
 \label{stdnf}
 \ba
 \Xi^{\rm TS}(\kappa, 0, 2 \pi)&=\exp \left[ {\log 2 \over 2}-{\log \ri \kappa \over 4}+\CA(\ri \kappa)  \right]   \vartheta_1\left( { \xi}( \ri \kappa)+{1\over 4},  { \tau}( \ri\kappa) \right),
\ea
 \ee
 where 
 \be \vartheta_1(v,\tau)= \sum_{n \in \IZ} (-1)^{n-1/2} \exp \left[ \pi \ri \left( n+1/2\right)^2 \tau + 2 \pi \ri \left( n +1/2\right) v\right].\ee

\subsection{Spectral theory and $q$-Painlev\'e  } \label{qpnp}

\begin{figure}[h!] \begin{center}
 {\includegraphics[scale=0.35]{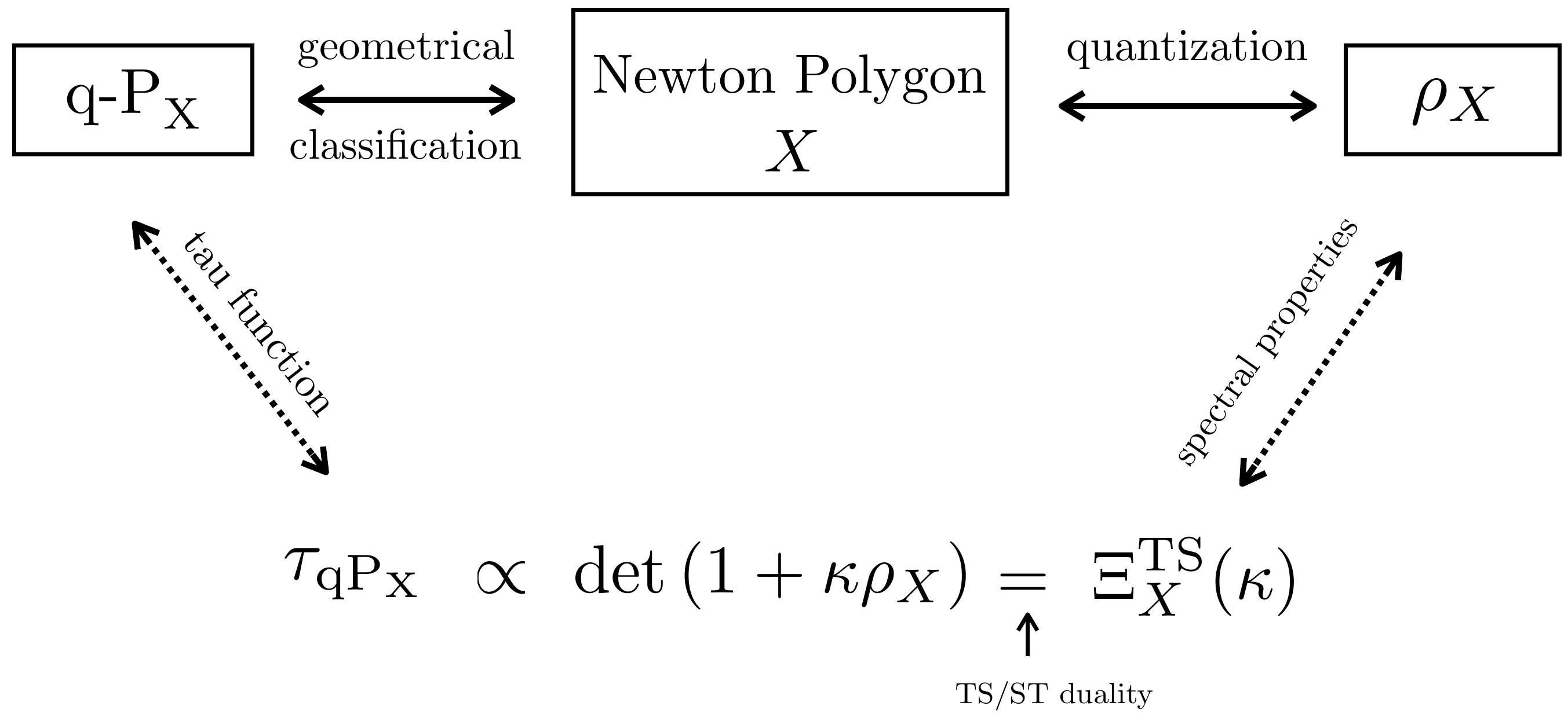}}
\caption{ A schematic representation of the correspondence between q-Painlev\'e equations, spectral theory and topological strings. We denoted by $\rho_X$ the operator \eqref{rhoo} and by $\Xi^{\rm TS}_X$ the  grand canonical partition function of topological string  \eqref{ts}.  }\label{schema}
\end{center}
\end{figure}

 In  the present  work we give a concrete link between $q$-Painlev\'e equations and the TS/ST duality of \cite{ghm}. 
  Our proposal is the following. As explained in \cite{yamadaameba,Ormerod:2014aa} one can associate a Newton polygon to a class of $q$-Painlev\'e equations in Sakai's classification (see \figref{p-toric}). Such  polygons represent the rational surfaces which are used to classify Painlev\'e equations in \cite{sakai}.    Once the Newton polygon of a given $q$-Painlev\'e   equation has been identified, we can apply the quantisation procedure presented in section \ref{s2}. We expect  the resulting Fredholm determinant  \eqref{Fred} to compute the $\tau$ function of the given $q$--Painlev\'e equation.  This is schematically represented on  \figref{schema}.
In this work we will test this proposal in detail for the   $q$--P$\rm III_3$, however it would be important to test this expectation for other $q$--Painlev\'e equations as well. 

\begin{figure}[h!] \begin{center}
 {\includegraphics[scale=0.5]{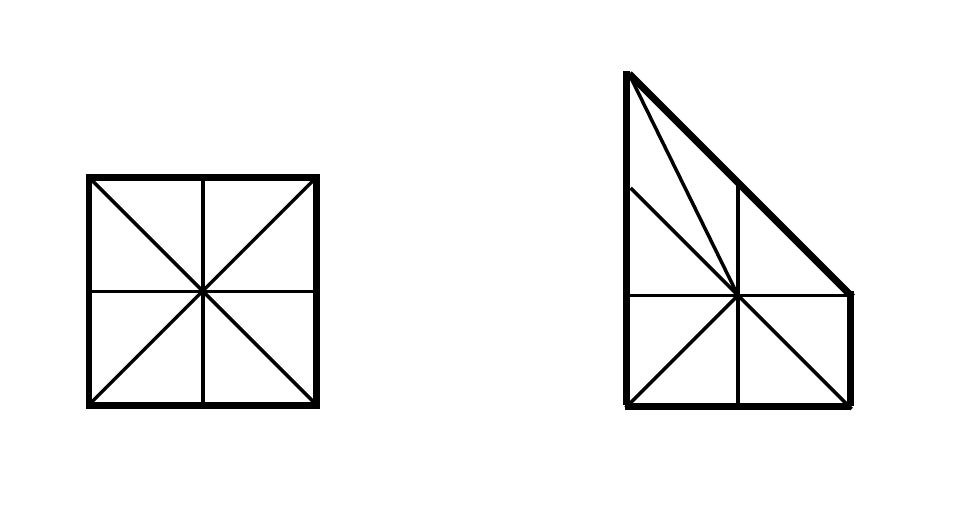}}
\caption{ Two  Newton polygons associated to $q$-PVI according to  \cite{yamadaameba, Jimbo1996}   }\label{seq2}
\end{center}
\end{figure}

We would like to observe that the correspondence  between $q$--Painlev\'e equations and Newton polygons is not always unique. Indeed given a  $q$--Painlev\'e equation, the idea of \cite{yamadaameba,Ormerod:2014aa} is that one can naturally associate to it a Newton polygon by looking  at the integral curves,  or conserved Hamiltonians, in the so-called autonomous limit. However, as discussed in  \cite{yamadaameba}, such integral curve admits different realisations. In turn, this is related to the fact that, as explained in \cite{Kajiwara:2015aa}, some of the $q$-Painlev\'e admit a classification in terms of both  blow ups of $\IP^1\times\IP^1$ or  blow ups of $\IP^2$.
For instance, in the case of  $q$-PVI  (surface type $A_2^{(1)}$ and symmetry  type $E_6^{(1)}$  in Sakai classification \cite{sakai}) there are two different ones
leading to the two different Newton polygons depicted in  \figref{seq2}.
Analogously, for the $q$-Painlev\'e $\rm III_3$ one can consider either  $\IP^1 \times \IP^1$ or  $\IF_2$ surfaces. 
Interestingly it was   pointed out in \cite{kmz,gkmr} that the spectral problems arising in the quantisation of the mirror curve to local $\IP^1 \times \IP^1$ and local $\IF_2$ are  equivalent. Therefore, upon a suitable normalisation and identification of the parameters, the  Fredholm determinants associated to these two manifolds are identified.  This is in perfect agreement with the proposal of this paper since they both compute the tau function of the same q-Painlev\'e equation. It would be interesting to see if the same identification holds also for other polygons  describing the same q-Painlev\'e equation such as for instance the ones in \figref{seq2}.  \footnote{ After the preliminary version of this work was sent to Yasuhiko Yamada, he proved the equality between the two spectral problems arising in the quantisation of  the two polygons  in \figref{seq2}. }

As an additional comment we notice that the self-dual point introduced in section \ref{sdpoint}    has a natural meaning from the $q$--Painlev\'e perspective. Indeed  the $q$--Painlev\'e equations can be studied in the so--called autonomous limit $q = 1$     \cite{qrt, qrt2, tsuda,Jimbo1996} in which they are expected to be solvable  by elliptic theta functions\footnote{ We thank Yasuhiko Yamada for discussions on this point.}.
From the TS/ST perspective $q=\re^{4 \pi^2 \ri /\hbar}$ and  $ q=1$ corresponds to
\be \label{sempl}\hbar={2 \pi \over m} \quad m \in \IN. \ee  It is possible to show   \cite{cgm, ghm, huang1606} that for these values the quantum theta function in \eqref{qtheta} becomes a conventional theta function.
 Moreover for $m=1,2$ the full spectral determinant can be expressed simply in terms of hypergeometric, Meijer and Jacobi theta functions as illustrated in section \ref{sdpoint}. Such  simplifications are expected from the  $q$--Painlev\'e viewpoint since  $q = 1$ correspond to the autonomous case.

\section{ Fredholm determinant solution for  $q$-Painlev\'e $\rm III_3$ } \label{s3}
Let us consider the canonical bundle over $\IP^1\times \IP^1$. Its mirror curve is given by \eqref{mirror} and the corresponding operator is
  \be \label{rhop1p1}\rho_{\IP^1 \times \IP^1}=\left( \re^{\mathsf x}+\re^{\mathsf p}+\re^{-\mathsf p}+m_{\IP^1 \times \IP^1}\re^{-\mathsf x}\right)^{-1}, \quad [\mathsf x,\mathsf p]=\ri \hbar. \ee
 The rescaled mass parameter $m$ 
 \eqref{rescaled} is given by
\be \log m_{\IP^1 \times \IP^1}={\hbar \over 2 \pi}\log m={\hbar \over 2 \pi} \xi .\ee 
Then according to the proposal of \cite{ghm} we have
\be \label{tsstp1}\det \left(1+\kappa \rho_{\IP^1 \times \IP^1}\right)=\sum_{n \in \IZ}\re^{\mathsf J_{{\IP^1 \times \IP^1}}(\mu+2 \pi \ri n, { \xi}, \hbar)}, \quad \kappa=\re^{\mu} \ee
where $\mathsf J_{{\IP^1 \times \IP^1}}$ is given in appendix \ref{def}. 
In this section by using the results of \cite{bsu} we show that, up to a normalisation factor, the term
\be \sum_{n \in \IZ}\re^{\mathsf J_{{\IP^1 \times \IP^1}}(\mu+2 \pi \ri n, { \xi}, \hbar)}\ee
 satisfies the 
$q$-deformed Painlev\'e ${\rm III_3}$  in the $\tau$ form.  As a consequence equation \eqref{tsstp1} provides a  Fredhom determinant representation for the $\tau$ function of $q$-P${\rm III_3}$. 
%I

%%
\subsection{The Bershtein--Shchechkin approach} 
It was observed in  \cite{bsu} that the $\tau$ function of  $q$-deformed Painlev\'e $\rm III_3$ can be written as
\be \label{qtau}\tau_{\rm C}(u,Z,q)= \sum_{n\in \IZ} s^n C(u q^{2n},q,Z) {\CZ (u  q^{2n}, Z,q^{-1},q)\over \left(uq^{2n+1},q,q\right)_{\infty} \left(u^{-1}q^{-2n+1},q,q\right)_{\infty}},\ee
where $\CZ $ are  the $SU(2)$ $q$-deformed conformal blocks at $c=1$. As explained in  appendix \ref{cbts} these correspond to the instanton partition function of topological string on local  $\IP^1 \times \IP^1$. Moreover $(\cdot, \cdot, \cdot)_{\infty}$ denotes the Pochhammer symbol and $C$ is a quite generic function.  The variable $s$ characterises the initial condition of the tau function and for the pourpouse of this paper we will always consider $s=1$. 
In \cite{bsu}  it was conjectured  and tested that $ \tau_{\rm C}(u,Z,q)$ satisfies the  $q$--deformed Painlev\'e ${\rm III_3}$ equation in the $\tau$ form, namely
\be \label{equa}Z^{1/4} \tau_{\rm C}(u,q,qZ) \tau_{\rm C}(u,q,q^{-1}Z)= \tau_{\rm C}(u,q,Z)^2+Z^{1/2}\tau_{\rm C}(uq,q,Z)\tau_C(uq^{-1},q,Z) \ee
 provided $C$ fulfils \cite{bsu} 
  \be\label{cond}\ba
{C(uq,q,Z)C(uq^{-1},q,Z)\over C(u,q,Z)^2}&=-Z^{1/2},\\
{C(uq,q,qZ)C(uq^{-1},q,q^{-1}Z)\over C(u,q,Z)^2}&=-uZ^{1/4},\\
{C(u,q,qZ)C(u,q,q^{-1}Z) \over C(u,q,Z)^2}&=Z^{-1/4}.
\ea\ee
The solution to these difference equations is clearly not unique and in \cite{bsu} several $C$-functions have been proposed.  In the next  section we  show that it is possible  to  chose $C$ in such a way that  $ \tau_{\rm C}(u,Z,q)$ is the spectral determinant of \eqref{rhop1p1} in the form conjectured in \cite{ghm}. This choice express $C$ in terms of $q$-deformed conformal blocks at $c=\infty$. 
Notice also that in \cite{bsu}, in order to ensure good convergence properties of \eqref{qtau} for a generic function $C$, they had to assume $\mid q \mid  \neq 1$. 
However,  as we will see in the following, by using an appropriate definition of $C$ one can make sense of \eqref{qtau} when $\mid q \mid  = 1$, which is the case studied in this paper \footnote{  On the operator side our result can be generalised straightforwardly  for any $\hbar$, however on the topological string side one has to be more careful to ensure the good convergence properties of $\Xi^{\rm TS}_X$ (see  \cite{gmprogress} for more details). }. This is important in order to study the autonomous limit of qP equations.

 \subsection{Tau function and quantum curve }
In this section we explain how to relate the results of \cite{bsu} to the TS/ST conjecture of \cite{ghm}.
As discussed previously the relevant geometry which makes contact with the  $q$-Painlev\'e $\rm III_3$ studied in \cite{bsu} is the canonical bundle over  $\IP^1 \times \IP^1$. In particular to connect \eqref{qtau} and  the r.h.s.~of \eqref{tsstp1} we use the following dictionary 
\be \label{dic}Z^{-1}=\re^{\xi},  \quad q=\re^{\ri 4 \pi^2/\hbar }, \quad  \quad u=\re^{\xi}Q_b, \quad Q_b=\re^{-{2\pi \over \hbar}t(\mu, \xi, \hbar)} ,\ee 
where $t(\mu, \xi,\hbar)$ denotes the quantum mirror map namely 
\be\label{mirrormap} \ba  t(\mu, \xi,\hbar)=& 2 \mu -2 (m_{\IP^1 \times \IP^1}+1) \kappa^{-2}+\kappa^{-4}\left(-3 m_{\IP^1 \times \IP^1}^2-\frac{2 m_{\IP^1 \times \IP^1} \left({ {\re }^{2\ri \hbar}}+4 {\re }^{\ri \hbar}+1\right)}{{\re }^{\ri \hbar}}-3\right)+O\left(\kappa^{-6}\right), \\
 & \kappa=\re^{ \mu}, \quad m_{\IP^1 \times \IP^1}=\re^{{\hbar \over 2 \pi}\xi }.\ea \ee
 When $\hbar$ is real, one can check numerically that the series \eqref{mirrormap} has a finite radius of convergence (see for instance \cite{hmmo}). For generic complex values of $\hbar$  one has to perform a partial resumation in $m_{\IP^1 \times \IP^1}$ but it is still possible to organise \eqref{mirrormap} into a convergent series as discussed in  \cite{gmprogress}. 
Notice that
\be t(\mu, \xi  \pm {4 \pi^2 \ri \over \hbar},\hbar)=t(\mu, \xi,\hbar).\ee
By using the dictionary \eqref{dic} we have that shifting
\be u\to q^n u, \quad n \in \IZ \ee
while leaving $Z$ invariant in \eqref{equa}, \eqref{cond} is equivalent to \be \mu \to \mu- \ri n \pi ,\quad  n \in \IZ ,\ee
in the language of section \ref{s2},
where we used
\be \label{tshift} t(\mu, \xi ,\hbar)+2n \pi\ri= t(\mu+n \ri \pi,\xi,\hbar).\ee
Likewise shifting
\be Z\to q^n Z, \quad n \in \IZ \ee
while leaving $u$ invariant is equivalent to \be \mu \to \mu- \ri n \pi,\quad \text{and} \quad \xi  \to \xi -4 n \ri\pi^2/\hbar    \quad n \in \IZ .\ee
Therefore by using the dictionary \eqref{dic} together with the above considerations we can write \eqref{equa} as
\be \label{taum} \ba \re^{-\xi/4} \tau_{\rm C}(\mu-\ri \pi,\hbar, \xi -{4 \ri\pi^2\over \hbar }) \tau_{\rm C}(\mu+\ri \pi,\hbar, \xi+ {4 \ri\pi^2\over \hbar })&=\tau_{\rm C}(\mu,\hbar,\xi)^2\\
&+ \re^{-\xi/2}\tau_{\rm C}(\mu+\ri \pi,\hbar,\xi)\tau_{\rm C}(\mu-\ri \pi,\hbar,\xi) .\ea \ee
In the rest of the paper we will use the notation $ \tau_C(\mu,\hbar,\xi)$ and   $\tau_C(u,q,Z)$ interchangeably. 
  We define $C_0(u ,q, Z)$ such that 
 \be \ba  \label{guess}{C_0(u ,q, Z)\CZ (u ,Z, q^{-1},q)\over \left(uq ;q,q\right)_{\infty} \left(u^{-1}q;q,q\right)_{\infty}}
 =&\re^{ \mathsf J_{\IP^1 \times \IP^1}(\mu, { \xi},\hbar)+{\mathsf J}_{{ \rm CS}}\left(\ri \pi +{1\over2} \xi,{2\pi^2 \over \hbar}\right)},\ea\ee
 where the variables on the two sides are related by the dictionary \eqref{dic}. Moreover $\mathsf J_{\IP^1 \times \IP^1}$ is the topological string grand potential \eqref{jts} for the canonical bundle over $\IP^1 \times \IP^1$ and we denoted by $J_{\rm CS}$ the non--perturbative Chern--Simons free energy, which can be identified with the topological string grand potential of the resolved conifold \cite{ho2} (see appendix \ref{def} for the full definition). 
  We also denote  \be  {Z_{\rm  CS}\left({\hbar}, \xi\right)}=\exp\left[J_{{ \rm CS}}\left({\ri \pi +{1\over2} \xi, {2\pi^2 \over \hbar}}\right)\right].\ee
 By using  the results of appendix \ref{shifts} it is easy to see that $C_0(u ,q, Z)$  defined as in \eqref{guess} fulfils  \eqref{cond}.
Therefore 
\be \label{inter} \tau_{{\rm C}_0}(\mu,\hbar,\xi)= {Z_{\rm  CS}\left({\hbar}, \xi\right)}\sum_{n \in \IZ}\re^{\mathsf J_{{\IP^1 \times \IP^1}}(\mu+2 \pi \ri n, { \xi}, \hbar)}\ee
satisfies the $q$- Painlev\'e $\rm III_3$ equation given in \eqref{taum}.
By using the conjectural expression for the spectral determinant given in \eqref{tsstp1}  together with  \eqref{inter}, it follows that
\be \label{tauf} \tau_{{\rm C}_0}(\mu,\hbar,\xi)= {Z_{\rm  CS}\left({\hbar}, \xi\right)} \det \left(1+\kappa \rho_{\IP^1 \times \IP^1}\right).\ee
Hence this choice of $C_0(u ,q, Z)$ provides a Fredholm determinant representation for the $\tau$ function of  the $q$-Painlev\'e $\rm III_3$ equation. This representation can be thought as a generalisation of \cite{zamo,wu1} for the $q$-difference equation since in the limit $\hbar \to \infty$ \eqref{tauf} reproduces the solution to differential Painlev\'e $\rm III_3$ presented in  \cite{zamo,wu1} (see section \ref{continu}).

 At this point the following question arises: can we prove directly that the r.h.s~of  \eqref{tauf} satisfy \eqref{taum} without using the TS/ST duality namely without using the expression of
\be \det \left(1+\kappa \rho_{\IP^1 \times \IP^1}\right)\ee in terms of enumerative invariants given in \eqref{tsstp1}?
Even though we do not have a proof of it,  in sections \ref{s5} and \ref{s6} we will test this
 by direct analytical an numerical computations.
\subsection{The self--dual point }
As we will see in section \ref{s4}, $ {Z_{\rm  CS}\left({\hbar}, \xi\right)}$ satisfy  \eqref{qal}.
In particular this means that we can replace \eqref{taum} with  a more explicit equation for \footnote{For sake of notation we omit the subscript $\IP^1\times \IP^1$ in $ \Xi^{\rm TS}$ } \be \Xi^{\rm TS}(\kappa, \xi,\hbar)= \sum_{n \in \IZ}\re^{\mathsf J_{{\IP^1 \times \IP^1}}(\mu+2 \pi \ri n, { \xi}, \hbar)}.\ee
More precisely  \eqref{taum} becomes
\be \label{xiots}\ba   & \Xi^{\rm TS}(-\kappa, \xi -{4 \ri\pi^2\over\hbar },\hbar)  \Xi^{\rm TS}(-\kappa, \xi+{4 \ri\pi^2\over\hbar },\hbar)(1+\re^{-\xi/2})= \Xi^{\rm TS}(\kappa,\xi,\hbar)^2+\re^{-\xi/2} \Xi^{\rm TS}( -\kappa,\xi, \hbar)^2\ea.\ee
At the self dual point $\hbar=2\pi$ and for $\xi=0$ this reads 
\be \label{xioabjb}  \ba & 2  \Xi^{\rm TS}(-\kappa , -2\pi \ri,2\pi) \Xi^{\rm TS} (- \kappa , 2\pi \ri, 2 \pi )= \Xi^{\rm TS}(\kappa ,0,2 \pi)^2+ \Xi^{\rm TS}( - \kappa,0,2 \pi)^2.\ea\ee 
By using the results of section \ref{sdpoint} it is easy to see that \eqref{xioabjb} becomes simply an identity between theta functions (we take ${\rm Re}(\kappa)>0$ )
\be\label{thetaidin}  \ba \vartheta_3\left( { \xi}(\ri \kappa), { \tau}(\ri \kappa) \right) \vartheta_4\left( { \xi}(\ri \kappa),  { \tau}(\ri \kappa)\right)=& \frac{ \ri}{\sqrt{\ri \kappa }}\left(\vartheta_1( \xi(\ri \kappa)+{1\over 4},  \tau(\ri \kappa)) 
\right)^2\\
&+\frac{\ri  }{\sqrt{\ri \kappa }}\left(\vartheta_1( \xi(\ri \kappa)-{1\over 4},  \tau(\ri \kappa))\right)^2,\ea\ee
where
\be \vartheta_4(v,\tau)=\sum_{n \in \IZ}(-1)^n\exp[{\ri \pi n^2 \tau+2\pi \ri n v}] \ee
while  the others quantities have been defined in section \ref{sdpoint}.  We also used
\be \Xi^{\rm TS}(-\kappa , -2\pi \ri,2\pi)= \Xi^{\rm TS}(\kappa , 2\pi \ri,2\pi) .\ee
Hence in this particular case the $q$-Painlev\'e equation leads to  %
 \be \label{thetaidnew}  \frac{\eta^8\left(4 \tau( \kappa)\right)}{\eta^8(\tau\left( \kappa)\right)}= {\kappa^2 \over 256}.\ee
where we used \be \eta(\tau)=\re^{\ri \pi \tau/12}\prod _{n=1}^{\infty } \left(1-\re^{2 \pi \ri n \tau}\right)\ee as well as the  following theta function identity  \footnote{We thank  Yasuhiko Yamada for bringing our attention on this identity.} 
\be\frac{\vartheta_1(v+1/4,\tau)^2+\vartheta_1(v-1/4,\tau)^2}
{4 \re^{\ri \pi \tau /2} \vartheta_3(v,\tau)\vartheta_4(v,\tau)}
=\prod_{n=1, n \notin 4\IN}^{\infty}(1-\re^{2 n\ri \pi \tau  } )^{-2} .\ee 
Let us look at \eqref{thetaidnew} as an equation for $\tau$. We can write it as as
\be \label{thetaidnew2} j(2\tau)=\frac{\left(\kappa ^4+16 \kappa ^2+256\right)^3}{\kappa ^4 \left(\kappa ^2+16\right)^2}\ee
where
\be j(2\tau)=\frac{\left(256 \Delta_\eta ^{16}+16 \Delta_\eta ^8+1\right)^3}{\Delta_\eta ^{16} \left(16 \Delta_\eta ^8+1\right)^2}, \quad  \Delta_\eta=  \frac{\eta \left(4 \tau( \kappa)\right)}{\eta(\tau\left( \kappa)\right)}. \ee
This is the well known expression for the j-invariant function as quotient of $\eta$ functions and can be easily derived by using the identities in appendix \ref{ideta}.
%It is easy to see that $ J(\tau)$ is modular invariant  and the small $\bar q=\re^{ 4\pi \ri \tau}$  expansion read
%\footnote{We thank Jie Gu for a discussion on this point} 
%\be J(\tau)=\frac{1}{\bar q}+744+196884 {\bar q}+21493760 {\bar q}^2+864299970 {\bar q}^3++20245856256 {\bar q}^4+\mathcal{O}({\bar q}^5),\ee
%
Hence  \eqref{thetaidnew2} is the well known relation between j-invariant and the modular parameter of the elliptic curve describing the mirror curve to local $\IP^1\times \IP^1$ whose solution is known to be \eqref{taudef}. \footnote{We thank Jie Gu for discussions on this point.}  Therefore in the self-dual case the q-Painlev\'e equation reduces 
to the well known relation   \eqref{thetaidnew2} which define the prepotential of the underling geometry. To capture all the gravitational corrections instead one should  consider $q$-Painleve  with $q \neq 1$.

\subsection{The $q$-deformed algebraic solution}\label{s4}
One of the immediate consequences of representing the $\tau$  function as a spectral determinant  is that one can easily  obtain the corresponding algebraic solution studied for instance in \cite{vg,Bershtein:2016uov}. 
In the case of  differential Painlev\'e ${\rm III}_3$  the $\tau$ function is characterised by the initial conditions \footnote{We follow the notation of \cite{ilt}.}  \be \left(\sigma, \eta\right) \ee
which correspond to the monodromy data of the related Fuchsian system.
 When $\eta=0$ the  $\tau$ function admits the following  spectral determinant  representation  \cite{bgt,zamo}
\be    \label{tau3d} \tau  \left(\sigma, T\right )=\re^{-\frac{\log (2)}{12} -3\zeta'(-1)}T^{1/16}\re^{-4 \sqrt{T}}  \det\left(1+{\cos(2 \pi \sigma)\over 2 \pi}\rho_{\rm 4D}\right)\ee
where 
 \be \label{opri} \rho_{\rm 4D}= \re^{-{4 T^{1/4} } \cosh(\mathsf x)}{4 \pi \over \left(\re^{\mathsf p/2}+ \re^{-\mathsf p/2} \right)}\re^{-{4 T^{1/4} } \cosh(\mathsf x)}, \qquad [\mathsf x, \mathsf p]= 2 \pi \ri  .\ee
 In particular when $\sigma=1/4$ we have 
 \be\label{taueasy4d} \tau(1/4, 0, T)=\re^{-\frac{\log (2)}{12} -3\zeta'(-1)}T^{1/16}\re^{-4 \sqrt{T}}.\ee
which reproduces the well known algebraic solution for Painlev\'e ${\rm III}_3$ \cite{vg, bsu}.

Similarly for the $q$-Painlev\'e equations the Fredholm determinant representation makes contact with  the $q$-analogue of the algebraic solution when $ \kappa =0$.  From \eqref{tauf} it follows that
\be \label{logt}\ba  \log \tau_{C_0}(\mu,\hbar, \xi) \mid_{\kappa=0}=& \log {Z_{\rm  CS}\left({\hbar}, {\xi}\right)}
.\ea \ee
Hence the $q$-difference Painlev\'e  $\rm III_3$ at $\kappa=0$ reads 
\be \label{qal} Z^{\rm CS}\left(\hbar,\xi + {4 \pi^2 \ri /\hbar}\right)Z^{\rm CS}\left(\hbar,\xi-{4 \pi^2 \ri /\hbar}\right)=Z^{\rm CS}\left(\hbar,\xi\right)^2\left( \re^{\xi/4}+\re^{-\xi/4}\right).\ee
By using \eqref{pertu}, \eqref{npcs} together with \eqref{sjp}, \eqref{snjp} it is easy to verify that
\eqref{qal} is indeed satisfied. As expected, up to an overall $\hbar$ dependent normalisation, the solution \eqref{logt} reproduces the q-deformed algebraic solution of \cite{bsu} provided we choose $C$ in \eqref{qtau} as in \eqref{guess}. 
 At the self--dual point $\hbar=2 \pi$  we have a particularly nice expression (we suppose $0<\re^{-\xi}<1$)  \eqref{maxsusy}
\be \label{zms} \ba  \log Z^{\rm CS}\left(2 \pi,\xi\right)=&\frac{\text{Li}_3(\re^{-\xi})}{8 \pi ^2}+\frac{\text{Li}_2(\re^{-\xi})\xi}{8 \pi ^2}-\frac{\xi^3}{96 \pi ^2}-\frac{\log (1-\re^{-\xi}) \xi^2}{16 \pi ^2}\\
&-\frac{\xi}{16}-\frac{1}{8} \log (1-\re^{-\xi})-\frac{1}{8} \log \left(\frac{\re^{-\xi/2}+1}{1-\re^{-\xi/2}}\right)+{A_c(4)\over 2},\ea\ee
where $A_c$ is defined in \eqref{aacc}.
It is easy to see that \eqref{zms} fulfils \eqref{qal}.
%Hence for  $0<\re^{-\xi}<1$ we have
%\be \ba \log Z^{\rm CS}\left(2 \pi,\xi {\pm 2\pi \ri }\right)=&\frac{\text{Li}_2(\re^{-\xi}) (\xi\pm 2\pi \ri )}{8 \pi ^2}-\frac{(\xi\pm2\pi \ri )^3 }{96 \pi ^2}-\frac{\log (1-\re^{-\xi})( \xi\pm 2\pi \ri )^2}{16 \pi ^2}\\
%&-\frac{\xi\pm 2\pi \ri }{16}-\frac{1}{8} \log (1-\re^{-\xi})-\frac{1}{8} \log \left(\frac{-\re^{-\xi/2}+1}{1+\re^{-\xi/2}}\right)+\frac{\text{Li}_3(\re^{-\xi})}{8 \pi ^2}+{A_c(4)\over 2}
%\ea \ee
 
 We  note that, to our knowledge, the generic Riemann--Hilbert  problem behind  $q$-deformed Painlev\'e  equations has not been found so far. However the algebraic solution to $q$-$\rm III_3$ is the grand potential of the resolved conifold as  defined in \cite{hmmo}. We observe that in  \cite{tb,Scalise:2017kjx}  the partition function of the resolved conifold  arises as a  tau function of a given Riemann--Hilbert  problem. 
 It would be interesting to see if  it exists a generalisation for the  Riemann--Hilbert  problem of  \cite{tb,Scalise:2017kjx}  whose tau function makes contact with the Fredholm determinants appearing in \cite{ghm} namely  \eqref{Fred},\eqref{tsst} and therefore with $q$-deformed Painlev\'e  equations.

%%%%%%%%%%%%%%%%%%%%%%%%%%

\subsection{The continuous limit}\label{continu}
We are now going to explain how to obtain the standard differential Painlev\'e ${\rm III}_3$ starting from the q-Painlev\'e   ${\rm III}_3$ written in the form \eqref{xiots}. It is more convenient to work with the variables
\be \xi, \quad Q_f=\re^{\xi} \re^{-{2 \pi \over \hbar}t(\mu , \xi, \hbar)}.\ee
Then we write \eqref{xiots} as
 \be\ba   \Xi(Q_f, \xi -{4 \ri\pi^2\over\hbar },\hbar)   \Xi(Q_f, \xi+{4 \ri\pi^2\over\hbar },\hbar)(1+\re^{-\xi/2})&=\Xi(Q_f,\xi,\hbar)^2\\
 &+\re^{-\xi/2} \Xi( \re^{4 \pi ^2 \ri /\hbar}Q_f,\xi, \hbar) \Xi( \re^{-4 \pi ^2 \ri /\hbar}Q_f,\xi, \hbar). ~~\\ 
 \ea\ee
 We omit the overscripts $\left. \cdot \right. ^ {\rm TS}$  or $\left. \cdot \right. ^ {\rm ST}$ because what follows holds both for  $ \Xi^{\rm TS}$ and for $ \Xi^{\rm ST}$.
 We write
\be\label{approx}\ba  &\Xi(Q_f, \xi -{4 \ri\pi^2\over\hbar },\hbar)  \Xi(Q_f, \xi+{4 \ri\pi^2\over\hbar },\hbar)=\Xi(Q_f, \xi,\hbar)^2 +\\
&\left({4 \pi ^2 \ri \over \hbar}\right)^2\left(- \left(\partial_\xi  \Xi(Q_f, \xi,\hbar) \right)^2+\Xi(Q_f, \xi,\hbar) \partial_\xi^2 \Xi(Q_f, \xi,\hbar) \right)+\mathcal{O}(\hbar^{-4}).\ea\ee
The continuous limit leading to differential  Painlev\'e ${\rm III}_3$   in the gauge theory language corresponds to the dual 4d limit introduced in \cite{bgt,bgt2}. Let us introduce a new set of variable $(T, a)$  such that
\be \xi= a \beta \epsilon-\log\left((4 \pi^2)^4\beta^4 T \epsilon^4\right), \quad Q_f=\re^{a \beta \epsilon}.\ee
and
\be \hbar={1\over \beta \epsilon}.\ee
Hence
\be \partial_\xi=-T\partial_T\ee
The continuous limit is obtained by sending
\be \beta \to 0.\ee
More precisely it was shown in \cite{bgt} that in this case one has
\be \Xi^{\rm TS }(\re^{\alpha 4 \pi ^2 \ri /\hbar}Q_f, \xi,\hbar) \quad  \xrightarrow{\beta \to 0}\quad  \Xi^{\rm TS }_{\rm 4d}(\sigma, T)=\re^{\frac{\log (2)}{12} +3\zeta'(-1)}T^{-1/16}\re^{4 \sqrt{T}} Z^{\rm NO}(\sigma+{\alpha \over 2}, T)\ee 
where $a=8 \pi^2 \ri \sigma$ and $Z^{\rm NO}(\sigma, T)$ is the Nekrasov--Okounkov partition function for pure $\CN=2$, $SU(2)$ gauge theory  in the selfdual $\Omega$ background. More precisely
\be \ba Z^{\rm NO}(\sigma, T) =& \sum_{n \in \IZ} {T^{(\sigma+n)^2} B(T, \sigma+n) \over  G(1-2(\sigma+n))G(1+2(\sigma+n))},\\
 B(T, \sigma)=&\left( 1+\frac{T}{2 \sigma ^2} +\frac{\left(8 \sigma ^2+1\right) T^2}{4 \sigma ^2 \left(1-4 \sigma ^2\right)^2}+\mathcal{O}(T^3)\right).
 \ea\ee
Likewise it was shown in \cite{bgt} that
\be \Xi^{\rm ST }(\re^{\alpha 4 \pi ^2 \ri /\hbar}Q_f, \xi,\hbar) \quad  \xrightarrow{\beta \to 0}\quad  \Xi^{\rm ST}_{\rm 4d}(\sigma, T)= \det\left(1+{\kappa(\sigma+{\alpha \over 2})}\rho_{\rm 4d}\right)\ee 
where \be  \kappa(\sigma)= {\cos(2 \pi \sigma)\over 2 \pi} \ee and $\rho_{\rm 4D}$ are defined in \eqref{opri}. In particular both  $\Xi^{\rm ST }$ and $\Xi^{\rm TS }$ are well defined in the limit $\beta \to 0$.
In this limit  \eqref{approx} reads
\be\label{approx2}\ba    \Xi_{\rm 4d}(\sigma, T)^2  +\left({4 \pi ^2 \ri \epsilon \beta }\right)^2\left(- \left(T \partial_T  \Xi_{\rm 4d}(\sigma, T)\right)^2+\Xi_{\rm 4d}(\sigma, T) (T \partial_T)^2 \Xi_{\rm 4d}(\sigma, T) \right)+ \mathcal{O}(\beta^{4}).\ea\ee
Hence \eqref{xiots} becomes
\be \label{almostp}\ba &16 \pi ^4 \sqrt{T} \epsilon ^2  \left( \Xi_{\rm 4d}(\sigma, T) ^2- \Xi_{\rm 4d}(\sigma+{1\over 2}, T) ^2-\Xi_{\rm 4d}(\sigma, T)  T^{3/2}\partial_T^2 \Xi_{\rm 4d}(\sigma, T) \right.\\
&\left.+T^{3/2} \left(\partial_T\Xi_{\rm 4d}(\sigma, T) \right)^2-\Xi _{\rm 4d}(\sigma, T) \sqrt{T}\partial_T \Xi_{\rm 4d}(\sigma, T) \right)+\mathcal{O}(\beta)=0\ea\ee
By keeping the leading order in $\beta$ and defining 
\be \label {tauxi4d}\tau(\sigma, T)=\re^{-4 \sqrt{T}}\Xi_{\rm 4d}(\sigma, T)\ee we can write \eqref{almostp} as 
\be \label{taufp} - \tau(\sigma, T)^2\left({T{\rd \over \rd T}}\right)^2\log \tau(\sigma,T)-  \sqrt{T} \tau(\sigma+{1\over 2}, T)^2=0,\ee
which is the Painlev\'e $\rm III_3$ in the $\tau$ form. More precisely equation \eqref{taufp} is also called  the Toda-like form of  Painlev\'e $\rm III_3$, see for instance \cite{Bershtein:2016uov}. Notice that if we take
\be \label{tau1} \tau(\sigma, T)=\re^{-4 \sqrt{T}}\Xi_{\rm 4d}^{\rm TS}(\sigma, T),\ee
then we recover the results of \cite{ilt} relating   Painlev\'e $\rm III_3$ to pure $SU(2)$ gauge theory. Likewise by setting
\be  \label{tau2} \tau(\sigma, T)=\re^{-4 \sqrt{T}}\Xi_{\rm 4d}^{\rm ST}(\sigma, T),\ee
we make contact with the  solution to  Painlev\'e $\rm III_3$ of \cite{zamo,wu1}. As explained in \cite{bgt} the expressions \eqref{tau1} and \eqref{tau2} are different representation of the same function, namely  the $\tau$ function of differential Painlev\'e $\rm III_3$.
 
\section{The $q$-deformed Painlev\'e  $\rm III_3$ and matrix models} \label{s5}
In this section we focus on the operator side of the TS/ST duality  and  we formulate the results of section \ref{s3}  by using the operator theory/matrix models point of view. 
We first recall that by using \eqref{qal} and \eqref{inter} we can write the $q$-deformed Painlev\'e  $\rm III_3$   \eqref{taum}  in the form \eqref{xiots}. This means that  at the level of the spectral determinant
\be \label{deet} \Xi^{\rm ST}(\kappa, \xi,\hbar)= \det \left(1+\kappa \rho_{\IP^1 \times \IP^1}\right)\ee
the $q$-deformed Painlev\'e  $\rm III_3$  equation reads
\be \label{xio}\ba   & \Xi^{\rm ST}(-\kappa, \xi -{4 \ri\pi^2\over\hbar },\hbar)  \Xi^{\rm ST}(-\kappa, \xi+{4 \ri\pi^2\over\hbar },\hbar)(1+\re^{-\xi/2})= \Xi^{\rm ST}(\kappa,\xi,\hbar)^2+\re^{-\xi/2} \Xi^{\rm ST}( -\kappa,\xi, \hbar)^2.\ea\ee
In the forthcoming sections we will perform several tests of  \eqref{xio}, even though a  proof is still missing.
\subsection{Matrix model representation}
By using standard results in Fredholm theory we express the  determinant  \eqref{deet}  in terms of  fermionic spectral traces 
\be \label{zst}Z(N,\hbar,\xi )= {1\over N!}\sum_{\sigma \in S_N}(-1)^\sigma \int {\rd ^N x}\prod_{i=1}^N \rho_{\IP^1\times \IP^1}(x_i, x_{\sigma(i)}),\ee
as%
\be \label{opt}  \Xi^{\rm ST}(\kappa, \hbar, \xi)=  \sum_{N\geq 0}\kappa^N Z(N,\hbar,\xi)\ee  where $S_N$ in \eqref{zst} is the group of permutations of $N$ elements and $\rho_{\IP^1\times \IP^1}(x,y)$ is the kernel of \eqref{rhop1p1}.
 Furthermore by using the Cauchy identity we can write \eqref{zst} as an $O(2)$ matrix model \cite{kmz}
 \be \label{matrixm}
Z(N,\hbar,\xi)={\re^{- \frac{ \hbar}{8 \pi } N \xi}\over N!}\int \frac{\rd^N z}{(2\pi)^N} \re^{- \sum_{i=1}^N (V(z_i,\hbar,\xi))} \frac{\prod_{i < j} (z_i-z_j)^2}{\prod_{i , j} (z_i+z_j)}, \ee
where the integral is over the positive real axis and
the potential is given by 
\be\label{fsb}\ba \re^{-V(z, \hbar,\xi)}
=& \re^{ \frac{b^2 \log {z}}{2}  }{\fad(\frac{b \log {z }}{2\pi}-{b\over 8 \pi}\xi + \ri b/4)\over \fad(\frac{b \log {z }}{2\pi}+{b\over 8 \pi}\xi - \ri b /4)}  
{ \fad(\frac{b \log {z }}{2\pi}+{b\over 8 \pi}\xi + \ri b /4)\over\fad(\frac{b \log {z }}{2\pi}-{b\over 8 \pi}\xi - \ri b/4)}
.
 \ea \ee
We use \be \hbar=\pi b^2 \ee and  $\fad$ denotes the Fadeev quantum dilogarithm  \cite{fk,faddeev}.
Our conventions for $\fad$  are as in \cite{kmz}.
Let us define 
\be \ba  {V_{\pm}(z, \hbar,\xi)}=&V(z, \hbar,\xi {\pm 4 \pi^2 \ri /\hbar}), \\
Z_{\pm}(N,\hbar,\xi)=&Z(N,\hbar,\xi {\pm 4 \pi^2 \ri /\hbar}). 
\ea\ee
Then we have
\be\label{mmpm} Z_{\pm}(N,\hbar,\xi)=\re^{- \frac{ \hbar}{8 \pi } N\xi} { (\mp \ri)^N\over N!}\int \frac{\rd^N z}{(2\pi)^N} \re^{- \sum_{i=1}^N (V_{\pm}(z_i,\hbar,\xi))} \frac{\prod_{i < j} (z_i-z_j)^2}{\prod_{i , j} (z_i+z_j)}.\ee
It follows that the $q$-difference Painlev\'e equation  \eqref{xio} is equivalent to the following relation between the matrix models  $Z_{\pm} (N,\hbar,\xi)$ and $Z(N,\hbar,\xi)$ 
\be \label{toshow} \ba &\sum_{N_1 =0 }^N Z(N_1, \hbar,\xi)Z(N-N_1, \hbar,\xi)\left( 1+\re^{-\xi/2}(-1)^N\right)\\
&= \sum_{N_1 = 0}^N  Z_+(N_1, \hbar,\xi)Z_-(N-N_1,\hbar,\xi)\left( (-1)^N+\re^{-\xi/2}(-1)^N\right).\ea\ee
Let us check this equation in the simplest example, namely $N=1$. We write \eqref{toshow} as
\be \label{1i}2(1- \re^{-\xi/2}) Z(1, \hbar,\xi)=-(1+\re^{-\xi/2})(Z_+(1, \hbar,\xi)+Z_-(1, \hbar,\xi))  .\ee
By using the properties of the quantum dilogarithm (see for instance appendix A in \cite{hel}) it is easy to see that\be \label{iduseful2} \ba
\re^{-V_-(z, \hbar,\xi)}=&\re^{-V_-(z^{-1}, \hbar,\xi)}\\
\re^{-V_{+}(z, \hbar,\xi)}=&- \re^{-V_{-}(z, \hbar,\xi)}\left(\frac{\left(\re^{\xi /4}+\ri z\right) \left(\re^{\xi /4} z+\ri \right)}{\left(\re^{\xi /4}-\ri  z\right) \left(\re^{\xi /4} z-\ri \right)}\right)\\
\re^{-V( \re^{\ri \pi /b^2}z, \hbar,\xi)}=&\frac{1-\ri \re^{\xi /4} z}{\re^{\xi /4} z-\ri }\re^{-V_{-}(z, \hbar,\xi)}.
 \ea\ee
Moreover  
\be \int_{\IR_+}{\rd z \over z}\re^{-V( \re^{\ri \pi /b^2}z, \hbar,\xi)}= \int_{\IR_+}{\rd z \over z}\re^{-V(z, \hbar,\xi)}, \qquad \xi, b \in \IR.\ee
Hence  \eqref{1i} is equivalent to
\be\label{jj} \int_{\IR_+}{\rd z \over z }H(z, \xi)\re^{-V_{-}(z, \hbar,\xi)}=0, \ee
where \be H(z,\xi)=\left(-\frac{2 e^{-\frac{\xi }{4}} \left(e^{\xi /2}-1\right) \left(z^2-1\right)}{\left(e^{\xi /4}-\ri z\right) \left(e^{\xi /4} z-\ri\right)}\right).\ee
Since  $ H(z,\xi)=- H(1/z,\xi)$ it follows that \eqref{jj} indeed holds.

 \subsection{Comment on  TBA } 
 In the previous sections we found that  the spectral determinant of the operator \eqref{rhop1p1} fulfils the $q$-deformed Painlev\'e ${\rm III_3}$ equation in the $\tau$ form.
  Moreover it is possible to show that such a  determinant is determined by a  TBA system \cite{zamo,tw}.
Let us review how this goes. We consider a trace class operator whose kernel is of the form
\be \label{twr}\rho(x,y)= {\re^{-u(x)-u(y)}\over 4 \pi \cosh\left({x-y\over 2}\right)}.\ee
Then the Fredholm determinant of  \eqref{twr}  is determined by a set of TBA equations whose explicit form can be found in  \cite{zamo}.
If we set \be \label{zamou} u(x)=t \cosh(x) \ee we can make contact with the solution \eqref{tau3d} of Painlev\'e ${\rm III}_3$ as explained in \cite{zamo}.
Let us take instead 
\be\label{myu} u(x)=- \log\left |f\left (\frac{b x}{2\pi} \right) \right|+{b^2\over 16} \xi \ee
where
\be f(x) =\re^{\pi x {b}/2}{\fad(x-{b\over 8 \pi}\xi+ \ri b/4)\over \fad(x+{b\over 8 \pi}\xi- \ri b /4)}, \qquad \hbar=\pi b^2 \ee
and $\fad$ denotes the  Faddeev's quantum dilogarithm.
With the choice \eqref{myu} the kernel \eqref{twr} is  related by unitary transformation to the one of \eqref{rhop1p1} (see \cite{kmz}) and therefore to $q-$P ${\rm III}_3$.
This connection with TBA system can be exploited to compute $Z(N,m,\hbar)$ in \eqref{matrixm} exactly for  finite values of $N,m,\hbar$. Some examples are provided in the next sections.

We also observe that in \cite{oz} the authors constructed a TBA system which determines the Fredholm determinant of the operator arising by quantizing the mirror to local $\IP^2$. It would be interesting to understand if one can systematically construct a TBA for each $q$-deformed  Painlev\'e equation.

  \subsection{The self--dual point } \label{sub5}
 
   At the self--dual point $\hbar=2\pi$ and for $\xi=0$ equation \eqref{toshow} gives a relation between the following $ O(2) $ matrix integrals:
\be \label{zn1} Z(N, 0,2\pi)= {1\over N!}\int {\rd^Nz\over (4 \pi)^N}\left( \prod_{i=1}^N{1\over (z_i+1)^2}\right) \prod_{i<j}\left(z_i-z_j \over z_i+z_j\right)^2\ee
\be  \label{zpm1}  Z_{\pm}(N, 0,2\pi) = {(\mp \ri)^N\over N!}\int {\rd^Nz\over (4 \pi)^N}\left( \prod_{i=1}^N{1\over (z_i^2+1)}\right) \prod_{i<j}\left(z_i-z_j \over z_i+z_j\right)^2.\ee
Both these matrix integrals have been evaluated exactly for various values of $N$ in \cite{hondao,hmo2} . For the first few values we have
\be   Z(1, 0, 2 \pi) =\frac{1}{4 \pi }, \quad Z(2, 0, 2 \pi) =\frac{1}{128} \left(1-\frac{8}{\pi ^2}\right), \quad Z(3, 0, 2 \pi) =\frac{5 \pi ^2-48}{4608 \pi ^3}, \ee
\be  Z_{+}(1, 0, 2 \pi) = -{\ri \over 8}, \quad Z_{+}(2, 0, 2 \pi) =- \frac{1}{32 \pi ^2}, \quad Z_{+}(3, 0, 2 \pi) =\ri \frac{10-\pi ^2}{512 \pi ^2},\ee
and similar expression can be obtained for higher $N$ as well. By using these exact values we checked that  \eqref{toshow} indeed holds for  $N=1, \cdots 13$.

\section{Connection with ABJ theory } \label{s6}

The $q$-deformed Painlev\'e written in the form \eqref{xio} is  similar to Wronskian like relations that have been found experimentally in ABJ theory \cite{ghmabjm}. 
 This link is not surprising since  topological string on local $\IP^1\times \IP^1$ and ABJ theory at level $k$ with gauge group $U(N) \times U(N+M)$ are closely related \cite{mpabjm, awhs,honda}. In order to connect these two theories one has to use the following dictionary \cite{hondao,matsumori}
\be \label{diz}\log m_{\IP^1 \times \IP^1}= \ri \hbar -2 \pi \ri M, \quad \hbar = \pi k.\ee
 Therefore shifting
 \be \xi  \quad \to \quad  \xi {\pm 4 \ri\pi^2/\hbar }  \ee 
 in topological string is equivalent to a shift of the rank of the gauge group in the ABJ  theory by
 \be M \to M \pm 1.\ee
 Let us denote the grand canonical partition function of ABJ theory by
 \be \Xi_{\rm ABJ}(\kappa, k, M)= \sum_{N\geq 0} \kappa ^N Z_{\rm ABJ}(N,M,k)\ee
 where $Z_{\rm ABJ}(N,M,k)$ is the partition function of ABJ theory at level $k$ and with gauge group $U(N)\times U(N+M)$. According to 
 \cite{awhs,honda,gm} we have
\be\label{zabj}
Z_{\rm ABJ}(N,M,k)= {1\over N!}  \int \prod_{i=1}^N {\rd x_i \over 4 \pi k} V_M(x_i) \prod_{i<j} \left( \tanh \left( {x_i - x_j \over 2 k } \right) \right)^2, 
\ee
where
\be
V_M (x)={1\over \re^{x/2} + (-1)^M \re^{-x/2}} \prod_{s=-{M-1\over 2}}^{M-1\over 2} \tanh {x+ 2 \pi \ri s\over 2 k}. 
\ee
It follows that \be\label{sdabj}  \Xi_{\rm ABJ}(\kappa, k, M)= \det\left(1+\kappa \rho_{\rm ABJ}\right)=\prod_{n \geq 0}\left(1+\kappa\re^{-E_n}\right) \ee
where 
\be  \label{rhoabj}\rho_{\rm ABJ}=\frac{1}{2 \cosh(\mathsf{v}/ 2)} \frac{1}{\re^{ \frac{\mathsf{u}}{2} }+(-1)^M \re^{-\frac{\mathsf{u}}{2} }}  \prod_{s=\frac{-M+1}{2}}^\frac{M-1}{2} \tanh \Big ( \frac{\mathsf{u} +2\pi \ri s}{2 k} \Big ) ,\qquad  [\mathsf{u}, \mathsf{v}]=2 \pi \ri k . 
\ee
This is a trace class operator acting on $L^2(\IR)$ and  we denote by $\re^{-E_n}$ its eigenvalues.  It was shown in \cite{kmz}, that by using the dictionary  \eqref{diz} we have
   \be  \label{rel}Z_{\rm ABJ}(N,M,k)=\re^{N \ri \pi k/4- N\ri \pi M/2}Z(N,\hbar,m).\ee
  \subsection{  Wronskian-like relations and $q$-Painlev\'e }
The $q$--Painlev\'e equation in the form
  \eqref{xio} and in the ABJ dictionary \eqref{diz} reads
\be \label{xioabj}  \ba &  \Xi_{\rm ABJ}(-\kappa \ri ,k,M+1)\Xi_{\rm ABJ}(\ri \kappa ,k,M-1)(1-\re^{2\pi \ri M/k})\\
 &=\Xi_{\rm ABJ}(\kappa ,k,M)^2-\re^{2\pi \ri M/k}\Xi_{\rm ABJ}( - \kappa,k,M)^2.\ea\ee 
 In the following we show that \eqref{xioabj} can be derived also by using the Wronskian-like relations of  \cite{ghmabjm}.   Let us recall the result of \cite{ghmabjm}.
 We factorise the  determinant \eqref{sdabj} according to the parity of the eigenvalues of $\rho_{\rm ABJ}$, namely
 \be \Xi_{\rm ABJ}(\kappa,k,M)=\Xi_+(\kappa,k,M)\Xi_-(\kappa,k,M),\ee
where \be   \Xi_+ (\kappa,k,M)= \prod_{n\geq 0}\left(1+\kappa \re^{-E_{2n}}\right), \quad \Xi_- (\kappa,k,M)= \prod_{n\geq 0}\left(1+\kappa \re^{-E_{2n+1}}\right).\ee
Then in \cite{ghmabjm} it was found experimentally \footnote{ By a detailed numerical analysis of the spectrum of the operator \eqref{rhoabj}.} that the following relations hold
\be \label{w1}
\ba
& \re^{\frac{M}{2k}\pi \ri} \Xi_+\left(\ri \kappa,k,M+1\right)\Xi_-\left(-\ri \kappa,k, M-1\right)\\
&\quad -\re^{-\frac{M}{2k}\pi \ri} \Xi_+\left(-\ri \kappa,k, M+1\right)\Xi_-\left(\ri \kappa,k, M-1\right)
=2\ri \sin\left( \frac{M\pi}{2k} \right) \Xi\left(\kappa,k, M\right) ,
\ea
\ee
and
\be \label{w2}
\ba
&\re^{-\frac{M}{2k}\pi \ri} \Xi_+\left(\ri \kappa,k,M-1\right)\Xi_-\left(-\ri \kappa,k, M+1\right)\\
&\quad +\re^{\frac{M}{2k}\pi \ri} \Xi_+\left(-\ri \kappa,k, M-1\right)\Xi_-\left(\ri \kappa,k, M+1\right)
=2\cos\left( \frac{M\pi}{2k} \right) \Xi\left(\kappa,k, M\right) .
\ea
\ee
These relations are similar to the Wronskian-like relations of  \cite{Bazhanov:1996dr}.
 Let us denote by
 \be W_1[\kappa]= \re^{\frac{M}{2k}\pi \ri} \Xi_+\left(\ri \kappa,k,M+1\right)\Xi_-\left(-\ri \kappa,k, M-1\right) -\re^{-\frac{M}{2k}\pi \ri} \Xi_+\left(-\ri \kappa,k, M+1\right)\Xi_-\left(\ri \kappa,k, M-1\right)\ee
 and 
 \be W_2[\kappa]= \re^{-\frac{M}{2k}\pi \ri} \Xi_+\left(\ri \kappa,k,M-1\right)\Xi_-\left(-\ri \kappa,k, M+1\right) +\re^{\frac{M}{2k}\pi \ri} \Xi_+\left(-\ri \kappa,k, M-1\right)\Xi_-\left(\ri \kappa,k, M+1\right).
\ee
By using these definitions it is easy to verify that
\be \label{l1}\ba & \frac{1}{2} \ri \csc \left(\frac{\pi M}{k}\right)\left(W_1[\kappa ]+\re^{\frac{\ri \pi M}{k}} W_1[-\kappa ]\right) \left(-W_2[\kappa ]+\re^{\frac{\ri  \pi M}{k}} W_2[-\kappa ]\right) \\
&=\left(1-\re^{\frac{2 \ri  \pi  M}{k}}\right) \Xi_{\rm ABJ} (-\ri \kappa,k ,M+1) \Xi_{\rm ABJ} (\ri \kappa,k ,M-1).\ea\ee
On the other hand by using  \eqref{w1} and \eqref{w2} we have
\be \label{l2}\ba & \frac{1}{2} \ri  \csc \left(\frac{\pi M}{k}\right)\left(W_1[\kappa ]+e^{\frac{\ri  \pi M}{k}} W_1[-\kappa ]\right) \left(-W_2[\kappa ]+e^{\frac{\ri  \pi M}{k}} W_2[-\kappa ]\right) \\
&= \Xi_{\rm ABJ} (\kappa,k ,M)^2-\re^{\frac{2 \ri  \pi  M}{k}} \Xi _{\rm ABJ}(-\kappa ,k,M)^2.\ea\ee
The combination of  \eqref{l1} with \eqref{l2}  leads to \eqref{xioabj}. Since \eqref{w1} and \eqref{w2} have been tested in detail  both numerically and analytically  (see \cite{ghmabjm}), this provides a further strong evidence for the conjecture that the Fredholm determinant of the operator  \eqref{rhop1p1} indeed computes the tau function of q-P$\rm{III}_3$.

\subsection{Additional tests}
 In this section we perform further tests of \eqref{toshow} by using several results obtained in the context of ABJ theory. By using \eqref{diz} and \eqref{rel} we write \eqref{toshow} as
 \be \label{toshowabj} \ba & \sum_{N_1 =0 }^N  \re^{-N\ri \pi k/4+ N\ri \pi M/2} Z_{\rm ABJ}(N_1, M,k)Z_{\rm ABJ}(N-N_1, M,k)\left( 1 -\re^{2 \pi \ri M/k}(-1)^N\right)=\\
 &  \sum_{N_1 = 0}^N    \re^{-N\ri \pi k/4+\ri \pi (-N+MN +2N_1)/2} Z_{\rm ABJ}(N_1, M+1,k)Z_{\rm ABJ}(N-N_1, M-1,k)(-1)^N\left(1-\re^{2 \pi \ri M/k}\right). \ea \ee
 This equation can be tested in detail for several values of $N, M, k$ thanks to the exact results for $Z_{\rm ABJ}(N,M,k)$ obtained in \cite{hondao,hmo2,py,ho} by using TBA like techniques. Let us illustrate this in one example. 
We consider \eqref{toshowabj}  for $M=1, k=3$. We have
    \be \label{toshowabj2} \ba & \sum_{N_1 =0 }^N  \re^{-N\ri \pi /4} Z_{\rm ABJ}(N_1, 1,3)Z_{\rm ABJ}(N-N_1,1,3)\left( 1 -\re^{2 \pi \ri /3}(-1)^N\right)=\\
 &  \sum_{N_1 = 0}^N  \re^{- N \ri \pi 3/4}(-1)^{N_1+N} Z_{\rm ABJ}(N_1,1,3)Z_{\rm ABJ}(N-N_1,0,3)\left(1-\re^{2 \pi \ri /3}\right). \ea \ee
where we used Seiberg-like duality of ABJ theory \cite{abj}  to set 
\be Z_{\rm ABJ}(N,2,3)=Z_{\rm ABJ}(N,1,3).\ee
The quantities 
 \be\label{zaa} Z_{\rm ABJ}(N,0,3), \quad Z_{\rm ABJ}(N,1,3),  \ee 
 have been computed exactly in \cite{hondao,hmo2} for $N=1, \cdots , 10$.
 For instance for the first few values of $N$ we have
 \be \ba Z_{\rm ABJ}(1,0,3)=&{1\over 12}, \quad Z_{\rm ABJ}(1,1,3)=\frac{1}{12} \left(2 \sqrt{3}-3\right),\\
 Z_{\rm ABJ}(2,0,3)=&\frac{\pi -3}{48 \pi }, \quad Z_{\rm ABJ}(2,1,3)=\frac{1}{432} \left(-27+14 \sqrt{3}+\frac{9}{\pi }\right).\\
  \ea \ee 
 The exact values of \eqref{zaa}  for higher $N$ can be found in \cite{hondao,hmo2,ho}.   By using these results  we have explicitly  checked that \eqref{toshowabj2} indeed holds for $N=1, \cdots 10$.
 Similar tests can be done for other values of $M,k$ providing in this way additional  evidence for \eqref{xio} and as a consequence for the conjecture that the Fredholm determinant of the operator \eqref{rhop1p1} computes the $\tau$ function of $q$-Painlev\'e $\rm III_3$.
 
\section{Conclusions and open questions}\label{s7}

In this work we conjecture that Fredholm determinants of operators associated to mirror curves on suitable Calabi-Yau backgrounds compute
$\tau$ functions of $q$--Painlev\'e equations. We test this proposal in detail for the  case of  $q$--Painlev\'e $\rm III_3$ which is related to topological strings on local $\IP^1 \times \IP^1$.  However, it would  be important to test, and eventually prove,  our proposal in the other cases, see  \figref{p-toric}.  

 At the self dual point $\hbar=2 \pi$  the spectral determinant reduces to a classical theta function up to a normalisation factor and the $q$-Painlev\'e equations to the well known relation between the   j-invariant and the modular parameter of the elliptic curve describing the mirror curve to local $\IP^1\times \IP^1$. Hence in this particular limit $q$-Painlev\'e equations determine the tree-level prepotential $F_0$ of the underling geometry.  It would be interesting to determine all the higher genus free energies $F_g$  explicitly starting from the  $q$-Painlev\'e at $q \neq 1$.
%
%
%
%The solution to $q$--Painlev\'e equation we propose is well defined also for $q=1$ and  can be evaluated at the so-called self-dual point of the TS/ST construction.
 %At this point the spectral determinant reduces to a classical theta function up to a normalisation factor and the $q$-Painlev\'e equations to a non-trivial relation among eta-functions on the mirror curve as displayed in (\ref{thetaidnew}). This simplified form of the spectral determinant suggests an {intriguing} relation with heat kernel equations arising from quantum background independence of topological string \cite{witten-bi}, which it would be very interesting to explore further.  
%
In addition it would be interesting to understand  better the relation between the self-dual point, which is determined  by $F_0$ and its derivatives, 
and the  autonomous limit of $q$-Painlev\'e equations or  QRT maps \cite{qrt, qrt2, tsuda}.
This would provide an interesting new link between supersymmetric gauge theories/topological strings and dynamical systems,
presenting the deformation of five dimensional Seiberg-Witten theory induced by a self-dual $\Omega$-background as a deformation of integrable mappings in two-dimensions, which appear in soliton theory and statistical systems \cite{qrt, qrt2}.

A special role in the tests that we perform is played  by ABJ  theory.  Actually, $q$--Painlev\'e $\rm III_3$ equation gives a relation between two ABJ theories  with different ranks of the gauge group.  In this case equation \eqref{xio} can be derived from the Wronskian-like relations of \cite{ghmabjm}.
It would be interesting to see if it is always possible to express the $q$--Painlev\'e equations  by using Wronskian-like relations. 
In the case of other $q$--Painlev\'e equations we expect the role of ABJ to be played by the other  superconformal  gauge theories which make contact with topological string on del Pezzo's surfaces as displayed in \figref{p-toric} (see for instance \cite{Moriyama:2017nbw, Moriyama:2017gye}).

Another related question is the connection with tt* equations. Indeed it is well known that differential Painlev\'e $\rm III_3$ arises  in the context of  2d tt* equations \cite{ttstar}. 
It would be interesting to see if and how the $q$-deformed case is related to the 3d tt* equations of \cite{Cecotti:2013mba}.  Likewise it would be interesting to understand if it exists a deformation of the 2d Ising model which makes contact with $q$-deformed Painlev\'e equation and the determinant \eqref{introdet}. 

A last  interesting point would be the generalisation to mirror curves with higher genus.  Indeed the conjecture of \cite{ghm} can be generalized to higher genus mirror curves by introducing the notion of generalised spectral determinant  \cite{cgm2}. However, the equations obeyed by these generalised determinants  both in five and in four dimensions are not known to us. Some first results in this direction are presented in \cite{bgt2} were a relation with 2d tt* geometry and Toda equation was found in a particular limit.
It would be interesting to see whether this continues to hold also in five dimensions at the level of relativistic Toda hierarchy.

\section*{Acknowledgements}
We would like to thank Mikhail Bershtein, Jie Gu, Oleg Lisovyy, Marcos Mari\~no,  Andrei Mironov,  Alexei Morozov and especially Yasuhiko Yamada for  many  useful discussions and correspondence. We are also thankful to Jie Gu for a careful reading of the manuscript. We thank Bernard Julia and the participants of the workshop "Exceptional and ubiquitous Painlev\'e equations for Physics" for useful discussions and the stimulating atmosphere. 
The work of G.B. is supported by the PRIN project "Non-perturbative Aspects Of Gauge Theories And Strings".
G.B. and A.G. acknowledge support by INFN Iniziativa Specifica ST\&FI.
A.T. acknowledges support by INFN Iniziativa Specifica GAST.

\appendix

 \section{The grand potential: definitions}
\label{def}
In this section we review the definition of the topological string grand potential $\mathsf J _X$ associated to a toric CY $X$ with genus one mirror curve. We mainly follow the notation of \cite{cgm2,Marino:2015nla}. As in section \ref{s2} we denote  \be \kappa=\re^{\mu} \ee  the "true" modulus of $X$,  
$\bf m_X$ the set of mass parameters,  $\bf m$ the rescaled mass parameters \eqref{rescaled} and
\be {\boldsymbol \xi}=\log {\bf m}.\ee
The  K\"ahler parameters of $X$  are denoted by $ t_i  $ and can be expressed in terms of the complex moduli through  the  mirror map:
\be t_i=c_i \mu +\sum_{j=1}^{r_X}a_{ij} \log m_X^{(j)}+\Pi(\kappa^{-1}, {\bf m}_X), \ee
where $\Pi$ is a series in $\kappa^{-1}$and ${\bf m}_X$ while $c_i, a_{ij}$ are constants which can be read from the toric data of the CY \cite{hkp,hkrs}. For instance for local $\IP^1\times \IP^1$ $c_2=c_2=2$, $a_{11}=0$, $a_{21}=-1$ . By using the quantum curve \eqref{op} one  promotes the  K\"ahler  parameters
to  quantum  K\"ahler  parameters  \cite{acdkv} which we  denote by \be  t_i ( \hbar)=c_i \mu +\sum_{j=1}^{r_X}a_{ij}  \log m_X^{(j)}+\Pi(\kappa^{-1}, {\bf m}_X, \hbar). \ee For instance when $X$ is the canonical bundle over $\IP^1\times \IP^1$ we have
\be t_1 ( \hbar)=t(\mu, \xi, \hbar), \quad t_2 ( \hbar)= t(\mu, \xi, \hbar)-{\hbar \over 2 \pi } \xi,\ee
where  (see also equation \eqref{mirrormap})
\be \label{tp1p1}\ba  t(\mu, \xi,\hbar)=& 2 \mu -2 (m_{\IP^1 \times \IP^1}+1) z+z^2 \left(-3 m_{\IP^1 \times \IP^1}^2-\frac{2 m \left({\re }^{2\ri \hbar}+4 {\re }^{\ri \hbar}+1\right)}{{\re }^{\ri \hbar}}-3\right)+O\left(z^3\right), \\
 & z=\re^{-2 \mu}, \quad m_{\IP^1 \times \IP^1}=\re^{{\hbar\over 2 \pi} \xi} .\ea \ee
 We introduce the topological string free energy
\be F^{\rm top}_{\rm X}\left({\bf t}, g_s\right)={1\over 6 g_s^2} a_{ijk} t_i t_j t_k +b_i t_i +F^{\rm GV}_X\left({\bf t} , g_s\right)\ee
with
\be
\label{Fgv}
F^{\rm GV}_X\left({\bf t}, g_s\right)=\sum_{g\ge 0} \sum_{\bf d} \sum_{w=1}^\infty {1\over w} n_g^{ {\bf d}} \left(2 \sin { w g_s \over 2} \right)^{2g-2} \re^{-w {\bf d} \cdot {\bf t}},
\ee
where $n_g^{ {\bf d}}$ are the Gopakumar--Vafa invariants of  $X$ and  $g_s$ is the string coupling constant. 
  The coefficients $a_{ijk}, b_i $ are determined by the classical data of $X$ . In the limit $g_s \to 0$ we have
\be F^{\rm top}_{\rm X}\left({\bf t}, g_s\right)\sim\sum_{g\geq 0} F_g({\bf t}) g_s^{2g-2}, \ee
where $F_g({\bf t}) $ are called the genus $g$ free energies of topological string. For instance we have
\be
\label{f01} \ba
F_0({\bf t})=&{1\over 6} a_{ijk} t_i t_j t_k  + \sum_{{\bf d}} N_0^{ {\bf d}} \re^{-{\bf d} \cdot {\bf t}},\\
F_1({\bf t})=&b_i t_i + \sum_{{\bf d}} N_1^{ {\bf d}} \re^{-{\bf d} \cdot {\bf t}},\ea
\ee
where $N_g^{\bf d}$ are the genus $g$ Gromov--Witten invariants.  When $X$ is a toric CY one has explicit expressions  for \eqref{f01} in terms of hypergeometric and standard functions (see for instance \cite{kz,mpabjm,hmmo}).
Let us discuss briefly the convergence properties of \eqref{Fgv}. Let us first note that  \eqref{Fgv} has poles for  $\pi^{-1}g_s \in \IQ$ which makes ill defined on the real $g_s$ axis. If instead $g_s \in \IC/\IR$, then \eqref{Fgv} diverges as a series in $\re^{- {\bf t}}$. Nevertheless by using instanton calculus  it is possible to partially resumm it and organise it into a convergent series \cite{bsu}.

Similarly we define the Nekrasov--Shatahsvili  free energy  as
\be
\label{NS-j}
F^{\rm NS}({\bf t}, \hbar) ={1\over 6 \hbar} a_{ijk} t_i t_j t_k +b^{\rm NS}_i t_i \hbar +\sum_{j_L, j_R} \sum_{w, {\bf d} } 
N^{{\bf d}}_{j_L, j_R}  \frac{\sin\frac{\hbar w}{2}(2j_L+1)\sin\frac{\hbar w}{2}(2j_R+1)}{2 w^2 \sin^3\frac{\hbar w}{2}} \re^{-w {\bf d}\cdot{\bf  t}},
\ee
where $N^{{\bf d}}_{j_L, j_R}$ are the refined BPS invariants of $X$.  Moreover
it exists a  constant vector ${\bf B} $, called the B-field , such that 
\be \label{B-prop}
N^{{\bf d}}_{j_L, j_R}\neq 0 \quad \leftrightarrow \quad (-1)^{2j_L + 2 j_R+1}= (-1)^{{\bf B} \cdot {\bf d}}.
\ee
 For local $ \IP^1 \times \IP^1 $ this can be set to zero \cite{hmmo,ghm}.  
 When $\hbar \to 0$ we recover the following genus expansion
\be\label{fgns} F^{\rm NS}({\bf t}, \hbar) =\sum_{g\geq 0} F_g^{\rm NS}({\bf t})\hbar^{2g-2}.\ee
The convergent properties for the NS free energy are analogous to the ones \eqref{Fgv}.

The topological string grand potential is defined as 
\be
\label{jtotal}
\mathsf{J}_{X}({\mu}, \boldsymbol{\xi},\hbar) = \mathsf{J}^{\rm WKB}_X ({\mu}, \boldsymbol{\xi},\hbar)+ \mathsf{J}^{\rm WS}_X 
({\mu},  \boldsymbol{\xi} , \hbar),
\ee
where 
\be
\label{jws}
\mathsf{J}^{\rm WS}_X({\mu}, \boldsymbol{\xi}, \hbar)=F^{\rm GV}_X\left( {2 \pi \over \hbar}{\bf t}(\hbar)+ \pi \ri {\bf B} , {4 \pi^2 \over \hbar} \right).
\ee
Moreover 
\be
\label{jm2}
\mathsf{J}^{\rm WKB}_X({\mu}, \boldsymbol{\xi}, \hbar)= {t_i(\hbar) \over 2 \pi}   {\partial F^{\rm NS}({\bf t}(\hbar), \hbar) \over \partial t_i} 
+{\hbar^2 \over 2 \pi} {\partial \over \partial \hbar} \left(  {F^{\rm NS}({\bf t}(\hbar), \hbar) \over \hbar} \right) + {2 \pi \over \hbar} b_i t_i(\hbar) + A({\boldsymbol \xi}, \hbar),
\ee
where  $ A({\boldsymbol \xi }, \hbar)$ denotes  the so--called constant map contribution. 
It is important to notice that, even tough both $\mathsf{J}^{\rm WS}_X$ and $\mathsf{J}^{\rm WKB}_X$ have a dense set of poles on the real $\hbar$ axis, their sum \eqref{jtotal} is free of poles. In particular $\mathsf{J}_{X}$ is well defined for any value of $\hbar$.
Moreover, we have
\be \mathsf{J}^{\rm WKB}_X({\mu}, \boldsymbol{\xi}, \hbar)= {1\over 12 \pi \hbar} a_{ijk} t_i(\hbar) t_j(\hbar) t_k(\hbar) + \left( {2 \pi b_i \over \hbar} + {\hbar b_i^{\rm NS} \over 2 \pi} \right) t_i(\hbar) + 
\CO\left( \re^{-t_i(\hbar)} \right)+ A({\boldsymbol \xi}, \hbar).
\ee
Hence it is convenient to split 
\be \mathsf{J}^{\rm WKB}_X({\mu}, \boldsymbol{\xi}, \hbar)= {\rm P}_X({{\mu}, \boldsymbol{\xi}, \hbar})+  \mathsf{J}^{\rm WKB, inst}_X({\mu}, \boldsymbol{\xi}, \hbar)+  A({\boldsymbol \xi}, \hbar)\ee
where $\rm P_X$ encodes the polynomial part in $t_i$ of  $ \mathsf{J}^{\rm WKB}_X$. 
 For local $ \IP^1 \times \IP^1 $ we have 
\be \mathsf {\rm P}_{\IP^1 \times \IP^1}(\mu, \xi,\hbar) = -\frac{\xi t(\mu,\xi ,\hbar )^2}{16 \pi ^2}+\frac{t(\mu ,\xi,\hbar )^3}{12 \pi  \hbar }-\frac{\hbar  t(\mu,\xi ,\hbar )}{24 \pi }+\frac{\pi  t(\mu,\xi ,\hbar )}{6 \hbar }-\frac{\xi}{24}.  \ee
The constant map contribution for  local $ \IP^1 \times \IP^1 $ reads \cite{yhum}
\be \label{cm} A( \xi,\hbar)= A_p(\xi,\hbar)-J_{\rm CS}\left({2\pi^2 \over \hbar},\ri \pi +{1\over2} \xi\right),\ee
where
\be   A_p(\xi,\hbar)={\hbar^2 \over (4 \pi^2)^2}\left[\frac{\xi^3}{24}+\frac{\pi^2  \xi}{6 }\right]+A_c\left({\hbar\over \pi}\right), \ee with 
\be \label{aacc}
A_{\rm c}(k)= \frac{2\zeta(3)}{\pi^2 k}\left(1-\frac{k^3}{16}\right)
+\frac{k^2}{\pi^2} \int_0^\infty \frac{x}{\re^{k x}-1}\log(1-\re^{-2x})\rd x.
\ee
Moreover $J_{\rm CS}(g_s,t)$ is the non--perturbative Chern--Simons free energy \cite{npcs1,ho2}. As explained in  \cite{ho2} this also coincides with the grand potential of the resolved conifold as defined in \cite{hmmo} namely
\be\label{jconi}\ba 
J_{\rm CS}(g_s,T+\ri \pi)=&-g_s^{-2}\frac{T^3}{12}-g_s^{-2}\frac{\pi ^2 T}{12}-{1\over 24}T+{1\over 2}A_c(4 \pi /g_s)+\sum_{n \geq 1}{1\over n}\left(2 \sin {n g_s \over 2}\right)^{-2}(-1)^n\re^{-n T}\\
&-\sum_{n\geq 1} {1\over 4 \pi n^2}\csc\left({2 \pi^2 n \over g_s}\right)\left({ 2 \pi n  \over g_s} T + {2 \pi^2 n \over g_s}\cot \left({2 \pi^2 n \over g_s}\right)+1\right)\re^{-{2 \pi n \over g_s}T}.
\ea\ee
We also denote  
\be \label{pertu}  \ba J^{\rm pert}_{{ \rm CS}}\left(g_s,T+\ri \pi\right)&=-g_s^{-2}\frac{T^3}{12}-g_s^{-2}\frac{\pi ^2 T}{12}-{1\over 24}T+{1\over 2}A_c(4 \pi /g_s)+\sum_{n \geq 1}{1\over n}\left(2 \sin {n g_s \over 2}\right)^{-2}(-1)^n\re^{-n T}, \ea\ee
 \be \label{npcs}J^{\rm np}_{{ \rm CS}}\left(g_s,T+\ri \pi\right)=-\sum_{n\geq 1} {1\over 4 \pi n^2}\csc\left({2 \pi^2 n \over g_s}\right)\left({ 2 \pi n  \over g_s} T + {2 \pi^2 n \over g_s}\cot \left({2 \pi^2 n \over g_s}\right)+1\right)\re^{-{2 \pi n \over g_s}T}.\ee
 For instance when $g_s=\pi$ and $T>0$ we have  \cite{ho2}
 \be\label{maxsusy} \ba
 J_{\rm CS}(\pi,T+\ri \pi)=&-(\pi)^{-2}\frac{T^3}{12}-\frac{T}{12}-{1\over 24}T+{1\over 2}A_c(4)+{1\over 8 \pi^2}{\rm Li}_3(\re^{-2T})+{T\over 4 \pi^2}{\rm Li}_2(\re^{-2T})\\
 &-\left({T^2\over 4 \pi^2}+{1\over 8}\right)\log \left(1-\re^{-2T}\right)-{1\over 4}{\rm arctanh}(\re^{-T}).
 \ea\ee

%%%
\section{Conformal blocks and topological strings}\label{cbts}
We follow the convention   of \cite{bsu}  and write the $c=1$ $q$-conformal blocks as
\be \label{cb} \ba
\CZ(u,Z,q_1,q_2)=&\sum_{\lambda,\mu} Z^{|\lambda|+|\mu|}\frac{1} {N_{\lambda,\lambda}(1,q_1,q_2)N_{\mu,\mu}(1,q_1,q_2)N_{\lambda,\mu}(u,q_1,q_2)N_{\mu,\lambda}(u^{-1},q_1,q_2)}
\ea\ee 
where the sum runs over all pairs $(\mu, \lambda)$ of Young diagrams. Moreover we use   
\be\label{Nlm}
 \ba
N_{\lambda,\mu}(u,q_1,q_2)
=&\prod_{s\in \lambda}
(1-u q_2^{-a_\mu(s)-1}q_1^{\ell_\lambda(s)}) \cdot 
\prod_{s \in \mu}
(1-u q_2^{a_\lambda(s)}q_1^{-\ell_\mu(s)-1})\ea \ee
where $a_{\lambda}(s), l_{\lambda}(s)$ are the arm length and the leg length of the box $s \in \lambda$. The leading order in $Z$ is obtain by considering the couples $(\square, 0)$ and  $(0,\square)$ which gives 
\be \CZ(u,Z,q,q^{-1})={Z \over (1-q)(1-q^{-1})}{2\over (1-u )(1-u^{-1} )}+ \mathcal{O}\left({Z u}\right)^2\ee
Hence by identifying 
\be {Z  u}= Q_B, \quad u=Q_F, \quad q=\re^{\ri g_s} \ee
we obtain 
\be  \CZ(u,Z,q,q^{-1})= {2 q  Q_B \over (1-Q_F)^2(1-q)^2}+\mathcal{O}\left(Q_B\right)^2 \ee
which corresponds to the instanton partition function  (without the one loop contribution) of topological string on local $\IP^1\times \IP^1$.
Therefore we have 
 \be  \exp \left[\mathsf{J}^{\rm WS}_{\IP^1 \times \IP^1}({\mu}, {\xi}, \hbar)\right]=  {{1 \over (Q_f q,q,q)_{\infty}^2}} \CZ (Q_f ,{Q_b\over Q_f}, \re^{4\pi^2 \ri /\hbar},\re^{-4\pi^2 \ri /\hbar}),\ee
 where \be Q_b=\re^{-{2 \pi \over \hbar}t (\mu, \xi, \hbar)}, \quad Q_f=\re^{\xi} Q_b  \ee
 and $t (\mu, \xi, \hbar)$ is defined in \eqref{tp1p1}.
\section{Some relevant shifts} \label{shifts}
We recall the dictionary \eqref{dic}  \be Q_b=\re^{-{2 \pi \over \hbar}t (\mu, \xi, \hbar)}, \quad Q_f=\re^{\xi} Q_b , \quad   \quad \xi= \log m={2 \pi \over \hbar}\log {m_{\IP^1\times \IP^1}}, \quad q=\re^{4 \pi^2 \ri /\hbar} \ee
\be t_1 ( \hbar)=t(\mu, \xi, \hbar), \quad t_2 ( \hbar)= t(\mu, \xi, \hbar)-{\hbar\over 2 \pi} \xi ,\ee
We define
\be F_1(\mu,\xi,\hbar)=\re^{ J_{\IP^1 \times \IP^1}^{\rm WKB, inst}(\mu, { \xi},\hbar)}.\ee
Then we have
\be \ba   & {F_1( \mu +\ri \pi,\xi,\hbar) F_1( \mu -\ri \pi,\xi,\hbar)\over F_1( \mu,\xi,\hbar)^2}= 1, \\
&  {F_1( \mu-2 \pi \ri , \xi   {-4  \ri\pi^2/\hbar },\hbar) F_1( \mu +2 \pi \ri,\xi+ {4  \ri\pi^2/\hbar },\hbar)\over F_1( \mu,m,\hbar)^2}= 1,\\
&   {F_1( \mu -\ri \pi, \xi   {-4  \ri\pi^2/\hbar },\hbar) F_1( \mu +\ri \pi,\xi +{4  \ri\pi^2/\hbar },\hbar)\over F_1( \mu,\xi,\hbar)^2}= 1 .
\ea\ee
Similarly we define
\be \ba F_2({\mu}, {\xi}, \hbar)=
&{(Q_f ^{-1}q,q,q)_{\infty} \over (Q_f q,q,q)_{\infty}}. \ea\ee
Then we have
\be \ba & {F_2( \mu +\ri \pi,\xi,\hbar) F_2( \mu -\ri \pi,\xi,\hbar)\over F_2( \mu,\xi,\hbar)^2}= -{1 \over Q_f } , \\
&  {F_2( \mu-2 \pi \ri , \xi  {-4  \ri\pi^2/\hbar },\hbar) F_2( \mu +2 \pi \ri,\xi+{4  \ri\pi^2/\hbar },\hbar)\over F_2( \mu,\xi,\hbar)^2}=  -{1 \over Q_f } ,\ea\ee
\be  {F_2( \mu -\ri \pi, \xi   {-4  \ri\pi^2/\hbar },\hbar) F_2( \mu +\ri \pi, \xi  +{4  \ri\pi^2/\hbar },\hbar)\over F_2(\mu, \hbar,\hbar)^2}=1,  \ee
where we used  \be  \left(u q;q,q\right)_{\infty} =\prod_{i,j\geq 0}\left(1-u q q^{i+j}\right)=\exp \left[- \sum_{s\geq 1}\frac{ u^s}{s \left(q^{s\over 2}-q^{-\frac{s}{2}}\right)^2} \right]. \ee
Similarly we define
\be F_3( \xi, \hbar)=\exp[A_p( \xi,\hbar) ]\ee
and we have
\be \ba  
&  {F_3(   \xi {-4  \ri\pi^2/\hbar },\hbar) F_3(  \xi+{4  \ri\pi^2/\hbar },\hbar)\over F_3( m,\hbar)^2}= \re^{-\xi/4}.
\ea\ee
Also for
\be F_4( \mu, \xi, \hbar)=\exp[\rm P_{\IP^1\times \IP^1}(\mu, \xi,\hbar) ]\ee
we have
\be \ba   & {F_4( \mu +\ri \pi,\xi,\hbar) F_4( \mu -\ri \pi,\xi,\hbar)\over F_4( \mu,\xi,\hbar)^2}=\re^{\xi/2} Q_b ,   \\
&  {F_4( \mu-2 \pi \ri , \xi   {-4  \ri\pi^2/\hbar },\hbar) F_4( \mu+2 \pi \ri ,\xi+{4  \ri\pi^2/\hbar },\hbar)\over F_4( \mu,\xi,\hbar)^2}= \re^{2 \xi} Q_b^2,
\\
&   {F_4( \mu -\ri \pi, \xi {-4  \ri\pi^2/\hbar },\hbar) F_4( \mu +\ri \pi,\xi+{4  \ri\pi^2/\hbar },\hbar)\over F_4( \mu,\xi,\hbar)^2}=\re^{\xi/2}.
\ea\ee
Moreover we have
\be \label{sjp}J^{\rm np}_{{ \rm CS}} \left({2\pi^2 \over \hbar},{1\over 2}\xi+ {2 \pi^2 \ri \over \hbar}+\ri \pi\right)+J^{\rm np}_{{ \rm CS}} \left({2\pi^2 \over \hbar},{1\over 2}\xi- {2 \pi^2 \ri \over \hbar}+\ri \pi\right)= 2J^{\rm np}_{{ \rm CS}} \left({2\pi^2 \over \hbar},{1\over 2}\xi+\ri \pi\right).\ee
Similarly
\be \label{snjp}\ba & J^{\rm pert}_{{ \rm CS}} \left({2\pi^2 \over \hbar},{1\over 2}\xi+ {2 \pi^2 \ri \over \hbar}+\ri \pi\right)+J^{\rm pert}_{{ \rm CS}} \left({2\pi^2 \over \hbar},{1\over 2}\xi- {2 \pi^2 \ri \over \hbar}+\ri \pi\right)- 2J^{\rm pert}_{{ \rm CS}} \left({2\pi^2 \over \hbar},{1\over 2}\xi+\ri \pi\right) \\
&={1\over 4}\xi+\log \left(\re^{-\xi/2}+1\right).\ea\ee
\section{Some identities for $\eta$ function}\label{ideta}
We denote
\be \eta(\tau)=\re^{\ri \pi \tau/12}\prod _{n=1}^{\infty } \left(1-\re^{2 \pi \ri n \tau}\right)\ee
the Dedekind $\eta$ function. The Weber modular functions are defined as 
\be\ba f(\tau)={\eta^2(\tau)\over \eta(\tau/2)\eta(2\tau)},\\
 f_1(\tau)={\eta(\tau/2)\over \eta(\tau)},\\
  f_2(\tau)=\sqrt{2}{\eta(2\tau)\over \eta(\tau)}.\\
\ea \ee 
Standard identities of  Weber modular functions are
\be f_1(\tau)^8+f_2(\tau)^8=f(\tau)^8,\ee
\be 8 j(\tau)=(f_1(\tau)^{16}+f_2(\tau)^{16}+f^{16}(\tau))^3,\ee
where $j$ is the j-invariant:
\be  j(\tau)={1\over {\bar q}}+ 744+196884{\bar q}+ \mathcal{O}({{\bar q}^2} ), \quad {\bar q}=\re^{2 \pi \ri \tau}\ee

\bibliographystyle{JHEP}
\bibliography{biblio}
\end{document}